\documentclass[twocolumn]{aastex63}
\usepackage{amssymb}
\usepackage{CJK}
\usepackage{perpage} 
\MakePerPage{footnote}
\usepackage[normalem]{ulem}
\graphicspath{ {figs/} }

\accepted{ by ApJ January4, 2024}
\shorttitle{AGNs in a MIR sample}
\shortauthors{Li et al.}

\begin{document}


\begin{CJK*}{UTF8}{gbsn}

\title{Active Galactic Nuclei in a Mid-Infrared Selected Galaxy Sample at $\mathbf{z>0.13}$: 

[\ion{Ne}{5}]$\lambda$3426 Line Emission as a Benchmark}

\author[0000-0001-7634-1547]{Zi-Jian Li}
\affiliation{Chinese Academy of Sciences South America Center for Astronomy (CASSACA), \\
National Astronomical Observatories of China (NAOC),\\
CAS, 20A Datun Road, Beijing 100012, China}
\affiliation{School of Astronomy and Space Sciences, University of Chinese Academy of Sciences, Beijing 100049, China}

\author[0000-0002-7928-416X]{Y. Sophia Dai}
\affiliation{Chinese Academy of Sciences South America Center for Astronomy (CASSACA), \\
National Astronomical Observatories of China (NAOC),\\
CAS, 20A Datun Road, Beijing 100012, China}
\correspondingauthor{Y. Sophia Dai}
\email{ydai@nao.cas.cn}

\author[0000-0001-6511-8745]{Jia-Sheng Huang}
\affiliation{Chinese Academy of Sciences South America Center for Astronomy (CASSACA), \\
National Astronomical Observatories of China (NAOC),\\
CAS, 20A Datun Road, Beijing 100012, China}
\affiliation{Harvard-Smithsonian Center for Astrophysics, 60 Garden Street, Cambridge, MA 02138, USA}

\author[0000-0003-3735-1931]{Stijn Wuyts}
\affiliation{Department of Physics, University of Bath, Claverton Down, Bath, BA2 7AY, UK}

\author[0000-0002-1335-6212]{Tian-Wen Cao}
\affiliation{Department of Astronomy, Xiamen University, 422 Siming South Road, Xiamen 361005, China}




\begin{abstract}

We present a 24$\mu$m-selected spectroscopic sample $z > 0.13$ (median $\langle z \rangle = 0.41$) in the Lockman Hole field, consisting of 4035 spectra. 
Our aim is to identify AGNs and determine their fraction in this mid-infrared selected sample. 
In this work, we use the [\ion{Ne}{5}]$\lambda$3426 emission line to spectroscopically identify AGNs.
Combined with broad-line Type I AGNs selected in our previous study, 
our sample consists of 887 ($\sim$22\%) spectroscopically confirmed AGNs. 
We perform a stacking analysis on the remaining spectra,
and find that in various
MIR-wedge-selected AGN candidates,
the stacked spectra still show significant [\ion{Ne}{5}]$\lambda$3426 emission,
In contrast, no clear [\ion{Ne}{5}]$\lambda$3426 signal is detected in non-AGN candidates falling outside the wedges. 
Assuming a range of AGN mid-IR SED slope of -0.3$<\alpha<$0.7,
and an average star-forming relation derived from 65 star-forming templates,
we develop a robust method to separate the AGN and star-forming contributions to the mid-IR SEDs
using the rest-frame L$_{12}$/L$_{1.6}$ vs L$_{4.5}$/L$_{1.6}$ diagram.  
We separate the objects into bins of L$_{12}$,
and find that AGN fraction increases with increasing L$_{12}$.
We also find that the stacked [\ion{Ne}{5}]$\lambda$3426 strength 
scales with L$_{12}$.
The pure AGN luminosity at 12$\mu$m exhibits
a positive correlation with the star formation rates, 
indicating possible co-evolution and common gas supply 
between the AGN and their host galaxies. 
Varying population properties across the redshift range explored contribute to the observed correlation.
\end{abstract}

\keywords{Active galactic nuclei --- Star formation }


\section{Introduction} \label{sec:intro}

Active galactic nuclei (AGN) play a crucial role in galaxy evolution, as demonstrated by previous studies highlighting strong correlations between central black hole mass and host properties \citep{2000ApJ...539L...9F,2000ApJ...539L..13G,2002ApJ...574..740T,2003ApJ...589L..21M,2004ApJ...604L..89H,2009ApJ...698..198G,2013ApJ...764..184M}.  
 It has been proposed that when a host experiences bulge growth, e.g. through merging, a central black hole will accrete rapidly \citep{2006ApJS..163....1H}.
Additionally, infrared observations have shown that AGNs are typically present in galaxies with intense star formation \citep{2003MNRAS.343..585F,2012A&A...545A..45R,2012A&A...540A.109S,2015ApJ...799...82C,2018MNRAS.478.4238D,2019ApJ...881..147S,2021ApJ...910..124X}. 
However, it is also believed that AGNs play a role in suppressing star formation \citep{2006MNRAS.365...11C,2010ApJ...717..708C,2012ARA&A..50..455F,2012Natur.485..213P,2021ApJ...923....6J}. 
Without this suppression, models predict more massive galaxies than observed \citep{2006MNRAS.370..645B}.
Thus, understand the relationship between AGN and star formation activities is critical to understanding the process of galaxy evolution.

Various methods for identifying AGNs in galaxies have been employed \citep{2017A&ARv..25....2P,2018ARA&A..56..625H,2022arXiv220906219L}.
The most common approach has been to use spectral signatures. AGNs can be classified into two types based on their spectra: Type I AGNs feature broad lines, whereas Type II AGNs have only narrow lines as their Broad Line Region (BLR) emission is presumed to be blocked by dust \citep{2001AJ....122..549V,2005AJ....129.1783H}.
Type II AGNs are more prevalent in galaxy samples \citep{2018ARA&A..56..625H}. 
\citet[hereafter BPT]{1981PASP...93....5B} proposed narrow line ratio diagnostics  to separate AGNs from star forming galaxies (e.g., [\ion{O}{3}]/H$\beta$ versus [\ion{N}{2}]/H$\alpha$). 
However, this BPT method \citep{2003MNRAS.346.1055K,2006MNRAS.372..961K} is only applicable for optical spectroscopic galaxy samples at z$<$0.4, when H$\alpha$ remains in the optical band.
Alternative methods using either rest-frame color or stellar mass, together with the [\ion{O}{3}]/H$\beta$ line ratio, have been developed to identify AGNs in galaxies at z$>$0.4 \citep{2011ApJ...728...38Y,2011ApJ...736..104J,2014ApJ...788...88J}, likewise exploiting high excitation lines as a marker of nuclear activity.
\citet{2003AJ....126.2125Z,2008AJ....136.2373R} constructed Type II quasar samples on the basis of their [\ion{O}{3}] luminosity, but this method is only effective for quasars with extremely strong [\ion{O}{3}] lines. 
Among more moderate luminosity systems, star formation can also excite [\ion{O}{3}] emission, thus compromising the approach \citep{2016MNRAS.462..181S}.
On the other hand, [\ion{Ne}{5}]$\lambda$3426 is a line that only appears in AGN spectra due to its high ionization potential, making it an attractive diagnostic tool for nuclear activity \citep{2013A&A...556A..29M,2022arXiv220906247C}.
Its shorter wavelength compared to [\ion{O}{3}]$\lambda$5007 implies it can be traced out to higher redshifts with optical spectrographs, at the expense of a greater sensitivity to dust obscuration \citep{2010A&A...519A..92G,2014A&A...571A..34V}.

In the absence of spectroscopic diagnostics, AGNs can also be identified on the basis of spectral energy distribution (SED) features.
In most infrared-selected galaxy samples, spectroscopic observations are expensive, and usually not feasible for every galaxy. 
To achieve this, several mid-infrared (MIR) color selections, also referred to as AGN wedges, were proposed  \citep{2004ApJS..154..166L,2005ApJ...631..163S,2012ApJ...748..142D,2013ApJ...772...26A}.
Though using different MIR colors, their selections in essence aim to isolate sources whose SEDs are dominated by the characteristic power-law emission of an AGN in the rest-frame MIR band \citep{2012ApJ...748..142D}.
These selections are usually applied to bright galaxy samples. 
Contamination by high-redshift galaxies becomes more severe if extended to fainter samples \citep{2010ApJ...713..970A}.
Despite their popularity, spectroscopic or X-ray observations remain the ultimate source of confirmation for AGNs.
Some spectroscopic observations on infrared-selected AGN candidates confirmed that about 80\% of their targets were AGNs, mainly of a dust obscured nature \citep{2007AJ....133..186L,2014ApJ...795..124H,2013ApJS..208...24L}.
Major observational campaigns with X-ray telescopes have also mapped the infrared survey fields \citep{2005MNRAS.356..568N,2009ApJS..185..433W,2015ApJS..220...10N,2016ApJ...819...62C,2017ApJS..228....2L}.
\citet{2015ApJS..220...10N} found that about 90\% of their X-ray sources have MIR counterparts.
\citet{2012ApJ...748..142D} further found that X-ray selected AGNs that predominantly reside in a small region within the broader Lacy AGN color wedge \citep{2004ApJS..154..166L}. 
Though X-rays represent a powerful technique to detect AGNs, some AGNs suffer from intrinsic extinction, and consequently can be absent from even the deepest X-ray observations \citep{2006ApJ...640..167A,2007ApJ...660..167D}.

In this paper, we present a spectroscopic study of a MIPS 24$\,\mu$m bright sample: 
these targets are selected from SWIRE survey \citep{2003PASP..115..897L} with F$_{24}> 400\mu$Jy in the Lockman Hole field.
We aim to provide a census of AGN in this infrared sample, which spans the redshift range of $z=0-6$, with $\sim$90\% of objects located below z=1.4.
[\ion{Ne}{5}]$\lambda$3426 emission serves as the primary AGN diagnostic in this study, where relevant augmented by BLR detections.
We also adopt the MIR AGN color selection designed by \citet{2004ApJS..154..166L} and \citet{2012ApJ...748..142D} for the dusty objects in our sample, and stack their spectra to search for the presence of [\ion{Ne}{5}]$\lambda$3426 as AGN confirmation.
We develop a method to isolate the AGN contribution to the SEDs of all objects in our spectroscopic sample, and study the relation between AGN and star formation activities.

The paper is organized as follows. 
In Section~\ref{sample}, we present the 24$\,\mu$m selected spectroscopic sample and associated multiwavelength data. 
The AGN selection methods are outlined in Section~\ref{AGN}.
Section~\ref{spectra} presents the composite spectra of the whole spectroscopic sample and the confirmation of AGN activity via stacking.
In Section~\ref{dec}, we present the SED decomposition into emission from AGN and star formation.
Section~\ref{pop} next presents the derived properties for all objects in our sample and discusses the relation between AGNs and their hosts. 
We summarize our conclusions in Section~\ref{sss}. 
Throughout the paper, we adopt a standard $\Lambda$CDM cosmology with $H_0 = 70\ {\rm km}\ {\rm s}^{-1}\ {\rm Mpc}^{-1}$, $\Omega_{M}$ = 0.3 and $\Omega_{\lambda}$ = 0.7. 
The initial mass function (IMF) used in this paper is from \citet{2001MNRAS.322..231K}.

\section{Sample and data} \label{sample}

\subsection{Spectroscopic sample}
\begin{figure}[ht!]
\epsscale{1.15}
\plotone{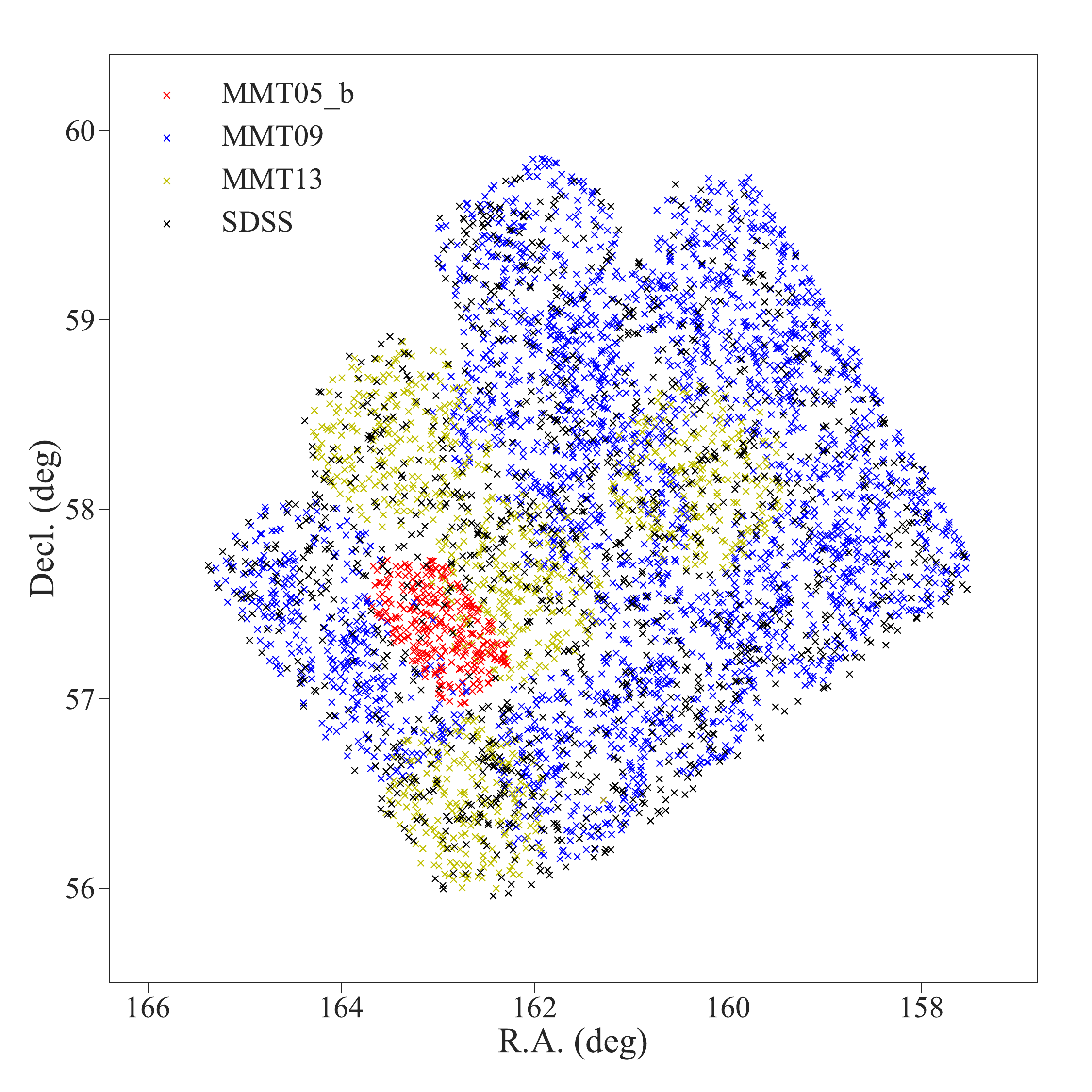}
\caption{Spatial distribution of spectroscopic coverage. The red, blue, yellow and black crosses are spectra observed in the MMT05$_{b}$, MMT09 and MMT13 runs, and as part of the SDSS survey, respectively.  The original MMT05 survey adopted a deeper 24$\mu$m limit:  F$_{24}> 60\mu$Jy. To ensure uniformity, we here keep the flux limit as F$_{24}> 400\mu$Jy and label this subset with subscript b. \label{ra_dec_fig}}

\end{figure}

\begin{figure*}[ht!]
\epsscale{1.15}
\plotone{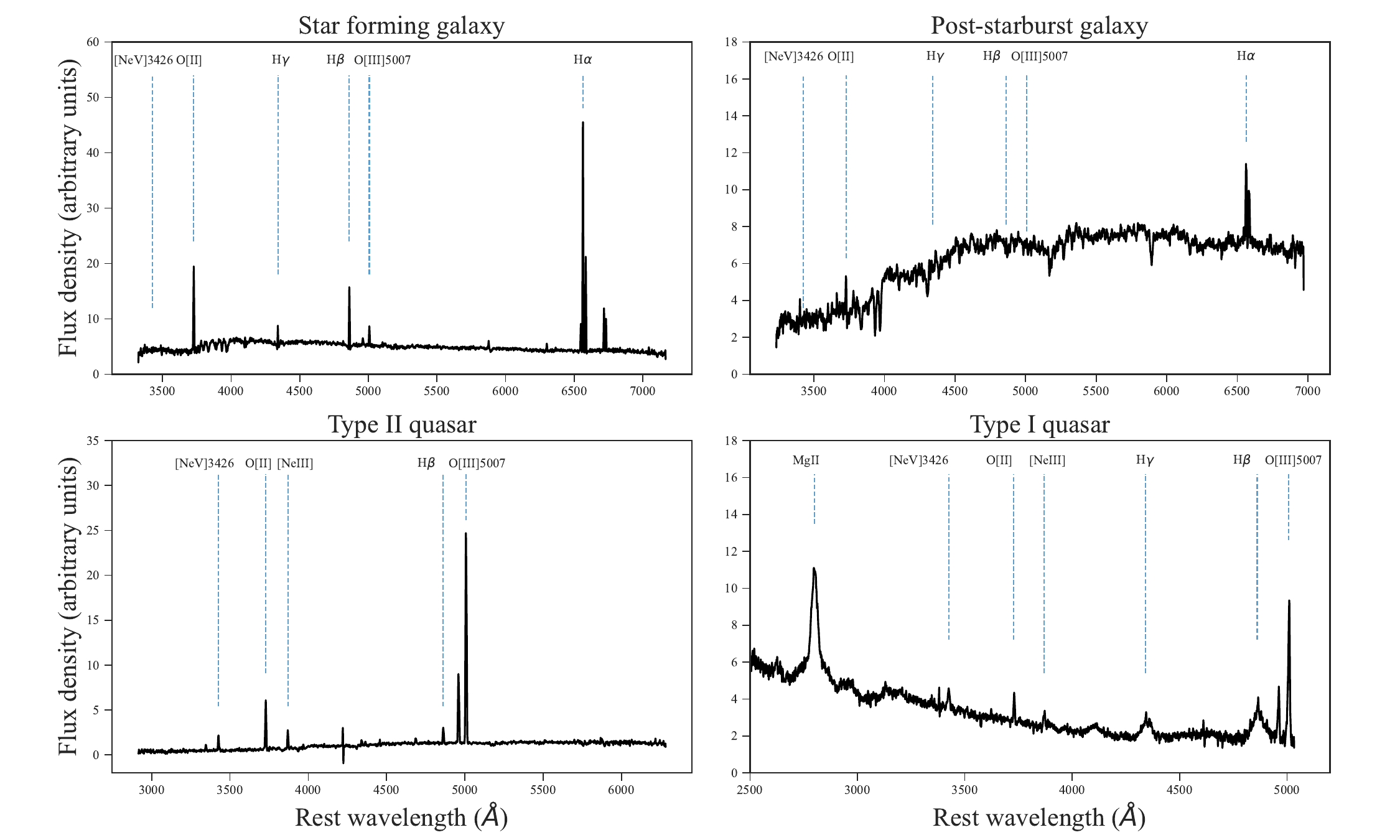}
\caption{Example spectra in rest-frame from our sample. \textit{Upper left panel}: Example of a star-forming galaxy. \textit{Upper right panel}: Example of a post-starburst galaxy. \textit{Lower left panel}: Example of a Type II quasar. \textit{Lower right panel}: Example of a Type I quasar.} \label{examp}
\end{figure*}

\citet{2014ApJ...791..113D} conducted a redshift survey on a MIPS 24$\mu$m selected sample.
The observations were done with Hectospec on MMT, an optical spectrograph (3850$\AA$-8300$\AA$) with 300 fibers deployed over a 1 deg$^{2}$ field of view.
The targets were chosen from the SWIRE survey \citep{2003PASP..115..897L} with F$_{24}> 400\mu$Jy and $r < 22.5$ mag.\footnote{The $r$ is the SDSS $r$-band model magnitude.}
There are 23,402 sources with F$_{24}> 400\mu$Jy within the 22 deg$^{2}$ area of the Lockman Hole field. Of these, 12,245 ($\sim$52\% of) sources are also brighter than $r=22.5$. 
In total, we configured 3,757 targets for observation, avoiding objects with existing SDSS redshifts in each Hectospec configuration.
\citet{2014ApJ...791..113D} presented the observing strategy for this redshift survey, in which they assigned each spectrum a redshift quality flag ranging from 1 to 4.
Robust redshifts were extracted for 3463 of the spectra (quality flags 3 and 4).
We further augmented the dataset with 1,317 spectra from SDSS DR17. 
Thus the total sample consists of 4780 spectra, a total of 4035 out of 4780 spectra have z $>$ 0.13.
Figure~\ref{ra_dec_fig} shows the layout of our survey in the Lockman-Hole field.
Example spectra for different spectral types are illustrates in Figure~\ref{examp}.

\begin{figure}[ht!]
\epsscale{1.15}
\plotone{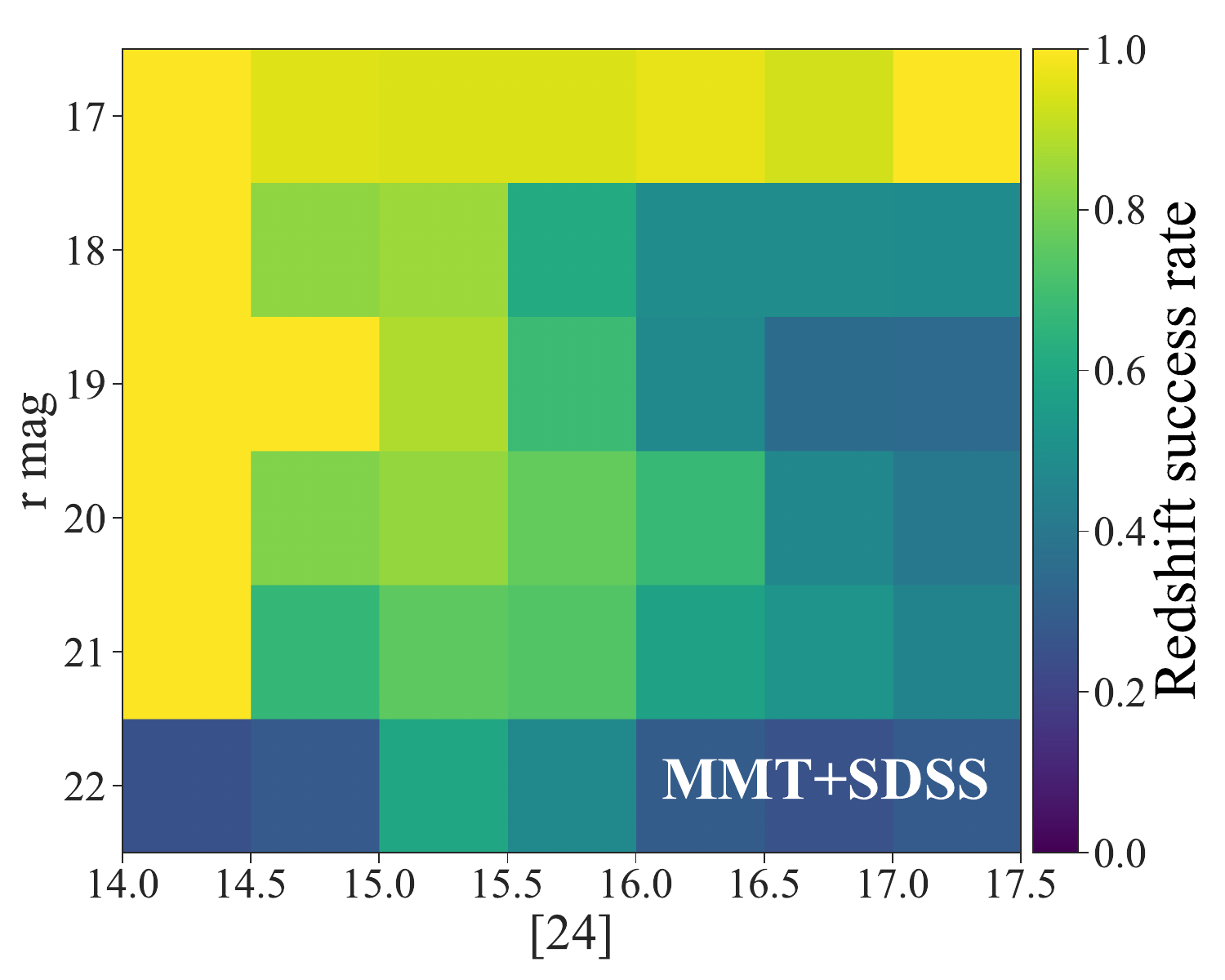}
\caption{Spectroscopic identification rate of our sample in the Lockman-Hole field. Redshifts used here are mainly from D14, with redshifts for the brightest objects in this field being supplemented from SDSS.
Each spectrum in \citet{2014ApJ...791..113D} was assigned a redshift quality flag from 4 to 1.
Spectra with quality flags 3 or 4 are considered to yield reliable redshifts, which are used for the statistics in this diagram. \label{completeness}}
\end{figure}

We present the success rate of obtaining spectroscopic redshifts for the 24$\mu$m selected sources in Figure~\ref{completeness}.
The identification rate remains approximately constant (and high) with varying 24$\mu$m flux at $r$-band magnitudes brighter than 17.7. 
The median value is 98\%. 
Below this limit, the success rate increases with increasing 24$\mu$m flux, from around 40\% at the faint end to about 70\% when the 24$\mu$m magnitude is brighter than 16.3 ($\sim$1000$\mu$Jy).
The median redshift success rate among all targets with $17.7 < r < 22.5$ is 54\%.

\subsection{Multiwavelength data}\label{sed_data}

We collected multiwavelength photometry for our spectroscopic sample from the rich ancillary data in the Lockman Hole field.  We cross-matched our sources with the Sloan Digital Sky Survey (SDSS) Data Release 17 (DR17) database, which provides photometric measurements (\texttt{cModelMag}) in the $u$, $g$, $r$, $i$, and $z$ bands \citep{2022ApJS..259...35A}. 
For the near-infrared (NIR), we derived $J$ and $K$ magnitudes within a $2"$ aperture using images from the UKIRT Infrared Deep Sky Survey (UKIDSS) DXS DR11PLUS \citep{2007MNRAS.379.1599L}.
The magnitudes were then aperture-corrected with a constant percentage appropriate for point sources.
In the MIR, we utilized aperture photometry provided by the SWIRE survey \citep{2003PASP..115..897L}, which covers a wavelength range of 3.6-24 $\mu$m. 
Finally, we obtained far-infrared (FIR) flux measurements from the Herschel Multi-tiered Extragalactic Survey (HerMES) in five bands spanning 100-500 $\mu$m \citep{2010MNRAS.409...48R}. 
The Herschel photometry was extracted using the XID algorithm using the positions of 24$\mu$m sources as priors \citep{2012MNRAS.424.1614O}.

We list the numbers and fractions of 3$\sigma$ detections in each band in Table~\ref{tab:data}.

\begin{deluxetable}{cccc}
\linespread{1}
\tabletypesize{\scriptsize}
\tablewidth{500pt} 
\tablenum{1}
\tablecaption{Multiwavelength data coverage. \label{tab:data}}
\tablehead{ \colhead{Band} 	& 	\colhead{\hspace{1cm}N$_{3\sigma}$\hspace{1cm}}	&	\colhead{\hspace{1cm}P$_{3\sigma}$}\hspace{1cm}} 

\startdata 
SDSS DR17  & 		&	\\
u  & \hspace{1cm}3402\hspace{1cm}  & \hspace{1cm}71\%\hspace{1cm}\\
g  & \hspace{1cm}4589\hspace{1cm}  & \hspace{1cm}96\%\hspace{1cm}\\
r  & \hspace{1cm}4751\hspace{1cm}  & \hspace{1cm}99\%\hspace{1cm}\\
i  & \hspace{1cm}4766\hspace{1cm}  & \hspace{1cm}99\%\hspace{1cm}\\
z  & \hspace{1cm}4446\hspace{1cm}  & \hspace{1cm}93\%\hspace{1cm}\\
UKIDSS DXS& 		&	\\
J  & \hspace{1cm}3587\hspace{1cm}  &\hspace{1cm}75\%\hspace{1cm}\\
K  & \hspace{1cm}3594\hspace{1cm}  &\hspace{1cm}75\%\hspace{1cm}\\
SWIRE & 		&	\\
3.6$\,\mu$m 	& \hspace{1cm}4396\hspace{1cm}  & \hspace{1cm}92\%\hspace{1cm}\\
4.5$\,\mu$m 	& \hspace{1cm}4396\hspace{1cm}  & \hspace{1cm}92\%\hspace{1cm}\\
5.8$\,\mu$m 	& \hspace{1cm}3940\hspace{1cm}  & \hspace{1cm}82\%\hspace{1cm}\\
8.0$\,\mu$m	    & \hspace{1cm}4049\hspace{1cm}  & \hspace{1cm}85\%\hspace{1cm}\\
24$\,\mu$m 	    & \hspace{1cm}4780\hspace{1cm}  & \hspace{1cm}100\%\hspace{1cm}\\
HerMES& 		&		&	\\
100$\,\mu$m  	& \hspace{1cm}1775\hspace{1cm}  & \hspace{1cm}37\%\hspace{1cm}\\
160$\,\mu$m  	& \hspace{1cm}1496\hspace{1cm}  & \hspace{1cm}31\%\hspace{1cm}\\
250$\,\mu$m  	& \hspace{1cm}2476\hspace{1cm}  & \hspace{1cm}52\%\hspace{1cm}\\
350$\,\mu$m  	& \hspace{1cm}1131\hspace{1cm}  & \hspace{1cm}24\%\hspace{1cm}\\
500$\,\mu$m  	& \hspace{1cm}347\hspace{1cm}  &  \hspace{1cm}7\%\hspace{1cm}\\
\enddata
\tablecomments{N$_{3\sigma}$ and P$_{3\sigma}$ are the numbers and percentages of 3$\sigma$ detections in each band, respectively.}
\end{deluxetable}

\section{AGN confirmation}\label{AGN}
Both AGN and star formation activities can contribute to the observed MIR emission. 
To study the connection between AGNs and their hosts, it is  crucial to firstly identify AGNs and estimate their contribution in a MIR-selected sample.
While the BPT diagram is a commonly used tool for identifying AGNs spectroscopically, its effectiveness is limited to low-redshift objects  \citep{2011ApJ...736..104J}, with only about 30\% of our spectra falling within a redshift range where BPT diagnostics are applicable. 
On the other hand, the [\ion{Ne}{5}]$\lambda$3426 line is considered a reliable AGN indicator \citep{2008ApJ...678..686A}.
Approximately 75\% of our spectra cover the rest-frame [\ion{Ne}{5}]$\lambda$3426 line region (i.e., those with a redshift between 0.13 and 1.40).
Here we note that [\ion{Ne}{5}] emission has doublet signals (peak at 3346 and 3426 $\AA$) in the optical regime. The 3346 line flux is usually one third of that of 3426 line \citep[e.g.,][]{2001AJ....122..549V}, thus throughout the paper we use the 3426 signal to confirm AGNs. 
The 3346 line is marked in the figures for display purpose but not used for analysis in this paper.
For the 487 spectra ($\sim$10\% of the whole sample) with $z>1.40$, we find that most of them ($\sim$92\%) are broad line AGNs. 
We refer the reader to \citet{2014ApJ...791..113D} for more details on this broad-line AGN sample and selection methods.
In the following section, we start by identifying AGNs with individual [\ion{Ne}{5}]$\lambda$3426 detections in the redshift range $0.13<z<1.40$. 

\subsection{[\ion{Ne}{5}]$\lambda$3426 selection}\label{NeV}

To identify [\ion{Ne}{5}]$\lambda$3426 emission, we calculate the equivalent width (EW) and signal-to-noise ratio (SNR) of the [\ion{Ne}{5}]$\lambda$3426 region for each spectrum.
We firstly define the line region as ($\lambda_{0}\pm13)\AA$. Here, $\lambda_{0}$ represents the line center defined as 

 \begin{equation}
\lambda_{0} =   \frac{ \sum_{n_{0}}  \lambda I_{\lambda}} {\sum_{n_{0}} I_{\lambda}},                       \label{eq0} 
\end{equation}

where $n_{0}$ is the number of pixels in the rest-frame wavelength range 3413-3439{\AA}. The 13{\AA} width is motivated by summing the typical [\ion{Ne}{5}]$\lambda$3426 line width (3$\sigma\sim$11{\AA}) and velocity shift of about 2{\AA} relative to [\ion{O}{2}] line center \citep{2001AJ....122..549V}.

 The SNRs are also calculated in the same line region using the following expression:
 
\begin{equation}
SNR = \frac{\sum_{n} I_{\lambda}-I_{c}}{\sqrt{n}\sigma_{c}}, \label{eq2} 
\end{equation}
 
where $n$ is the number of pixels in the line region, I$_{c}$ is the continuum intensity, I$_{\lambda}$ is the spectral intensity and $\sigma_{c}$ is the standard deviation of the continuum-subtracted intensity in the continuum region. 
We initially select 94 objects based on SNR $>$ 3. 
To ensure the fidellity of our sample, we conduct visual inspections to remove any fake detections caused by incorrect telluric line subtraction. 
Finally, we identify a total of 88 [\ion{Ne}{5}]$\lambda$3426-selected AGNs, with 43 of them being narrow-line Type II AGNs.

 \begin{figure}[ht!]
 \epsscale{1.1}
\plotone{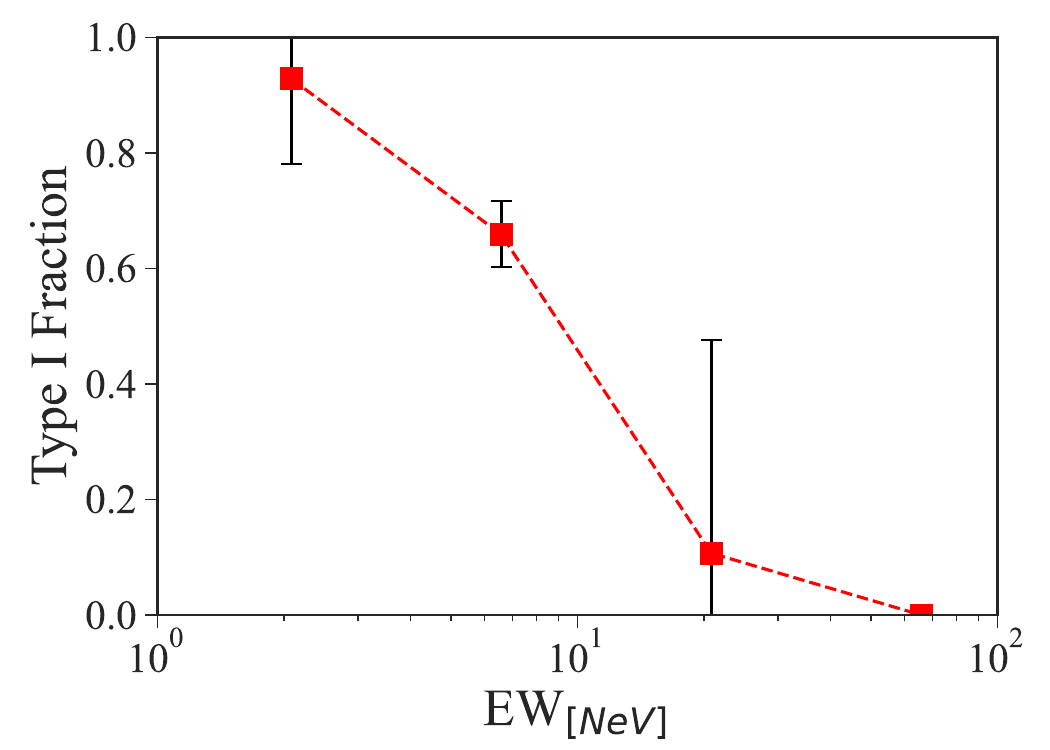}
\caption{Type I AGN fraction versus the equivalent width (EW) of [\ion{Ne}{5}]$\lambda$3426 for our [\ion{Ne}{5}]$\lambda$3426 selected AGN sample. The error bars show the Poisson uncertainty. \label{ew}}
\end{figure}


\subsection{The equivalent width bias}\label{bias}

Previous studies have suggested that the equivalent width (EW) of [\ion{Ne}{5}]$\lambda$3426 could be related to the degree of nuclear extinction \citep{2021ApJ...914...83Y}. 
In Type II AGNs, the torus obscures the continuum emission from the central region along the line of sight, leading to an increase of the EW of the emission line from the outer narrow-line region (NLR) \citep{2002ApJ...573L..81L}. 

We plot the Type I fraction among our [\ion{Ne}{5}]$\lambda$3426 selected sample as a function of EW in Figure~\ref{ew}. 
The fraction is defined as the ratio between the number of Type I AGN and the total number of AGN.
We find that the fraction decreases from 90\% to 10\% when the EW increases from $\sim$\,2{\AA} to 20\,{\AA}.
This result quantitatively illustrates the EW bias.


\subsection{The spectroscopically confirmed AGN sample}

\begin{figure}[ht!]
\epsscale{1.2}
\plotone{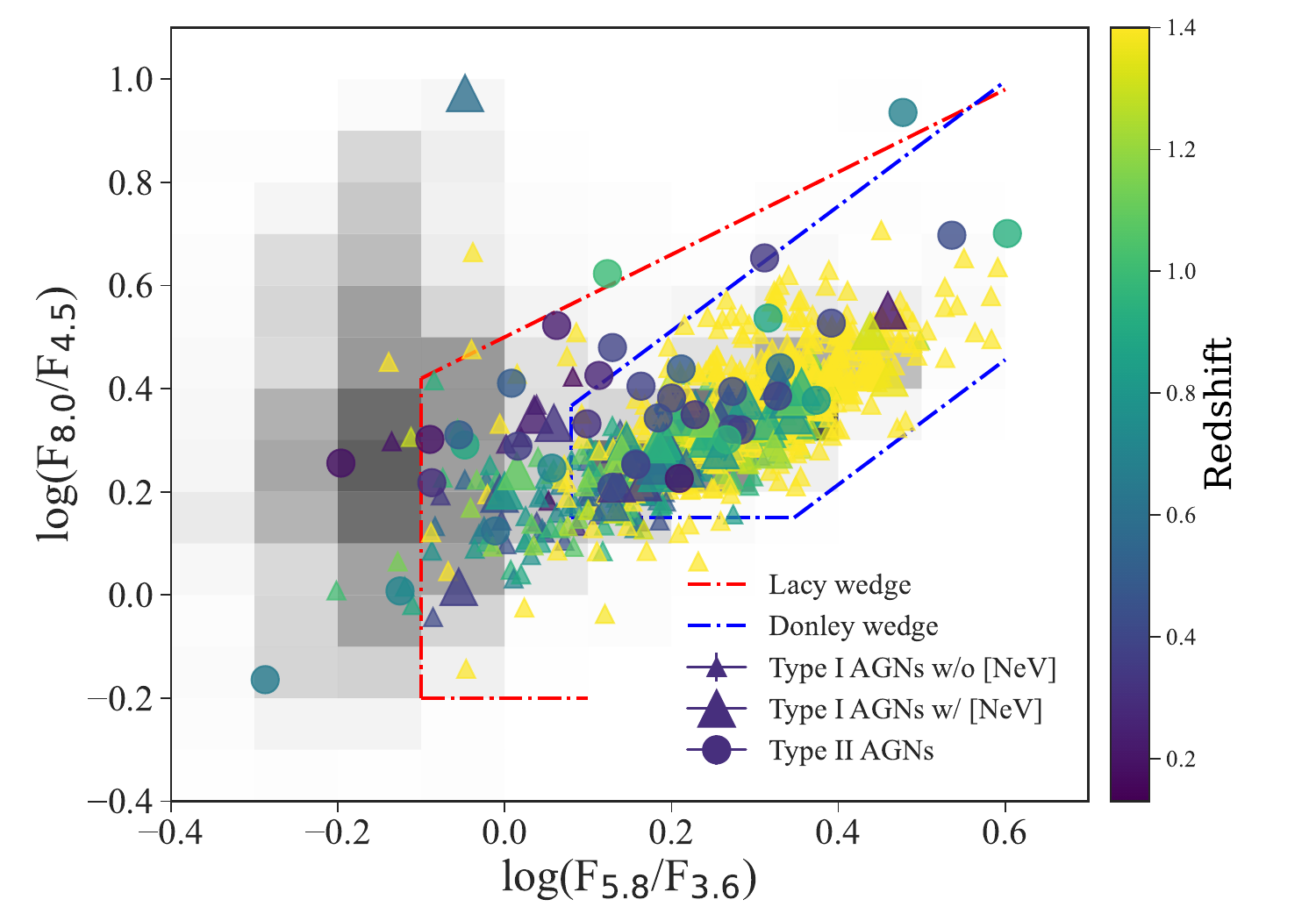}
\caption{Observed-frame color-color diagram of our spectroscopically confirmed AGNs. The greyscales denote the 24$\,\mu$m bright spectroscopic sample with secure redshifts. Triangle symbols mark the 844 Type I AGNs (A total of 45 of them have [\ion{Ne}{5}]$\lambda$3426 detections and these 45 objects are marked as triangles with larger size.) and circles mark the 43 Type II AGNs.  Colors indicate their redshifts.} The red and blue dash-dotted lines indicate the Lacy and Donley wedges, respectively. \label{lacy}
\end{figure}
The sample of broad-line AGN among our parent dataset was compiled by \citet{2014ApJ...791..113D}.
The combination of the [\ion{Ne}{5}]$\lambda$3426 and broad-line selections now provides a more complete picture of the AGN population.
This results in a total of 887 sources in our spectroscopically confirmed AGN sample. 
A total of 43 objects are [\ion{Ne}{5}]$\lambda$3426 selected Type II AGNs and the remaining 844 objects objects are Type I from \citet{2014ApJ...791..113D}. Among these Type I AGNs, there are 45 objects that also have [\ion{Ne}{5}]$\lambda$3426 detections.

Compared with MIR color selections of AGN, our optical spectroscopy confirms that 53\% and 84\% of AGN candidates\footnote{Hereafter we regard the objects selected by MIR color criteria but not confirmed by their spectra as AGN candidates.} selected by Lacy's and Donley's criteria, respectively, are true AGNs. As shown in Figure~\ref{lacy}, the Donley wedge is more successful in selecting strong AGNs at high redshifts. 

\section{Optical spectra of MIR-selected sample}\label{spectra}

\begin{figure*}[ht!]
\epsscale{1.15}
\plotone{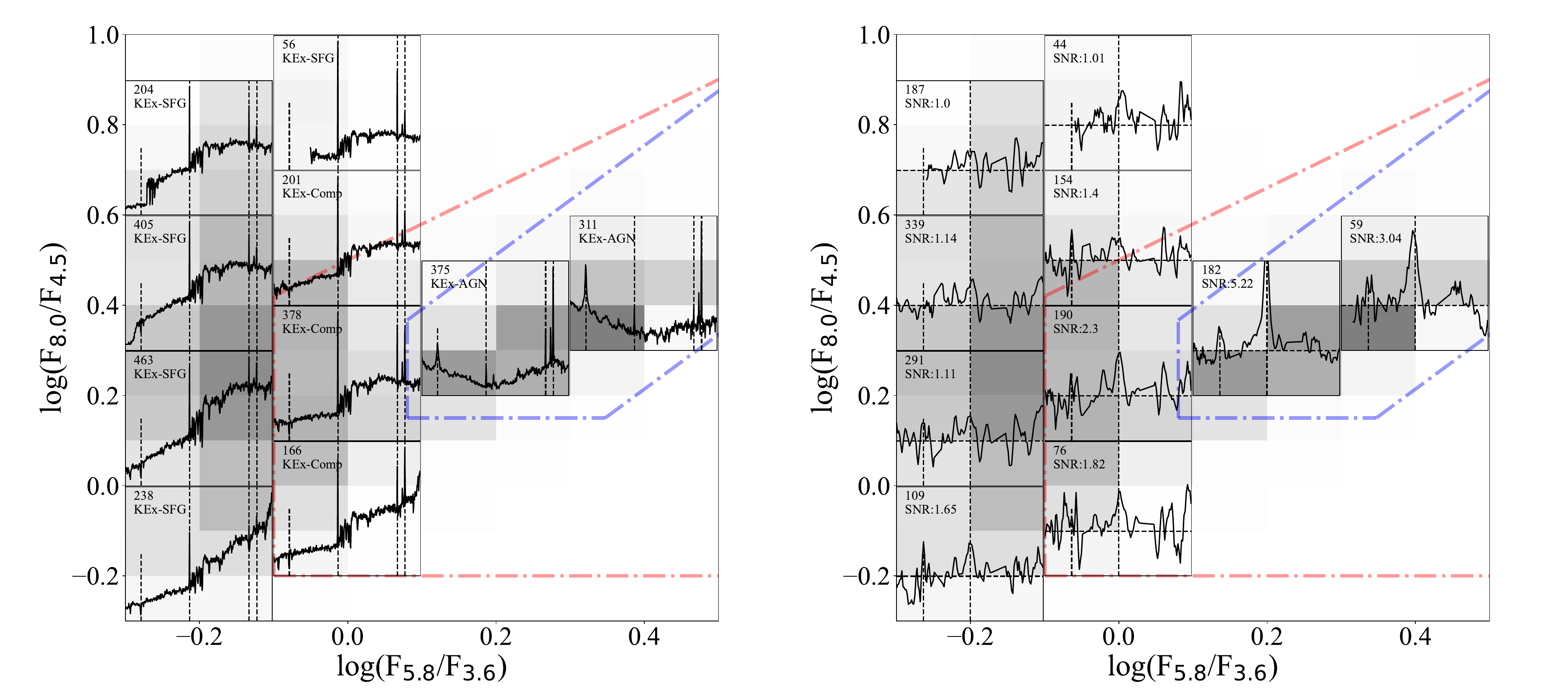}
\caption{\textit{Left panel}: Median composite spectra for objects binned by their location in the MIR color-color diagram. No redshift limit is applied in this panel. The rest-wavelength range of every inset panel is 2500-5200$\AA$. The number of combined spectra is marked. Besides, we mark KEx classification of each spectra in the upper left corner of each inset panel. The dashed lines mark \ion{Mg}{2}$\lambda$2800, [\ion{O}{2}]$\lambda$3727, H$\beta$ and [\ion{O}{3}$\lambda$5007 spectral lines from left to right, respectively. \textit{Right panel}: Mean stacking of the [\ion{Ne}{5}]$\lambda$3426 line region. In this panel, we only included spectra with $0.13<z<1.40$. The rest-wavelength range of every inset panel is 3300-3550$\AA$. Two vertical dashed lines show the locations of the [\ion{Ne}{5}]$\lambda$3346,3426 doublet. The number of spectra used in the stacking and the SNR of stacked [\ion{Ne}{5}]$\lambda$3426 is marked in the upper left corner of each inset panel.
The background greyshades in both panels depict the distribution of the whole 24$\,\mu$m bright spectroscopic sample. The red and blue dash-dotted lines represent the Lacy and Donley wedges, respectively. \label{lacy_stacking}}
\end{figure*}

\subsection{Composite spectra}\label{composite}
In this section, we present the composite spectra of objects in our spectroscopic sample and search for potential AGN signals across the MIR color space. 
To this end, we divide our sample into 10 subsets based on their position in the MIR color-color diagram (see Figure~\ref{lacy_stacking}). 
Each bin covers 0.2 dex in $\log(F_{5.8}/F_{3.6})$ and 0.3 dex in $\log(F_{8.0}/F_{4.5})$.  Where relevant, we label each bin by the coordinates of the lower-left corner (xmin, ymin).
To build the composite spectrum of each bin, we firstly shift all spectra to the rest frame and then normalize them to a common continuum level. 
Given the spread in redshifts within and between bins, the continuum level used for normalization is determined as the median flux in the rest-frame 2100-2200$\AA$ range for the (0.3,0.3) bin, the rest-frame  3100-3200$\AA$ range for the (0.1,0.2) bin, and the rest-frame  4200-4300$\AA$ range for the remaining bins. 
A common wavelength grid from 2500 to 5200$\AA$ was built, with wavelength steps of 6$\AA$.
In each color bin, we then 
construct the composite spectra by taking the median value of all the normalized data points included.
The results are shown in the left panel of Figure~\ref{lacy_stacking}. 
We can observe clear broad \ion{Mg}{2} line features in the two bins within the Donley wedge (i.e., the (0.1,0.2) and (0.3,0.3) bins ). 
They also show narrow H$\beta$ and strong [\ion{O}{3}] lines. This is because coverage of those lines requires relatively low redshifts, where in our sample the AGN family is in numbers dominated by Type II AGNs.
The spectrum in the (0.1,0.2) bin shows weaker and narrower \ion{Mg}{2} line compared to the spectrum in the (0.3,0.3) bin. 
Additionally, it displays a flater blue continuum \citep{2003AJ....126.2125Z}. 
\citet{2018ApJ...856..171Z} have develop the kinematics–excitation (KEx) diagram, which uses [\ion{O}{3}]$\lambda$5007/$H\beta$ line ratio and velocity dispersion of [\ion{O}{3}]$\lambda$5007 line, to classify the ionization source. 
We also label the KEx classification for each composite spectrum in the inset panel. 
We find that Donley objects are more like AGNs while the Lacy objects tend to be like composite galaxies.
The classification is also found to be consistent with the SNR of stacked [\ion{Ne}{5}]$\lambda$3426.
However, the [\ion{Ne}{5}]$\lambda$3426 stacking in the (-0.1,-0.2) bin shows an insufficient signal-to-noise ratio, which could be attributed to the smaller number of spectra available in that bin. 
Details on the [\ion{Ne}{5}]$\lambda$3426 stacking procedure and further discussions are provided in Section~\ref{sta}.

\subsection{AGN signals in color selected candidates}\label{sta}

\begin{figure}[ht!]
\epsscale{1.15}
\plotone{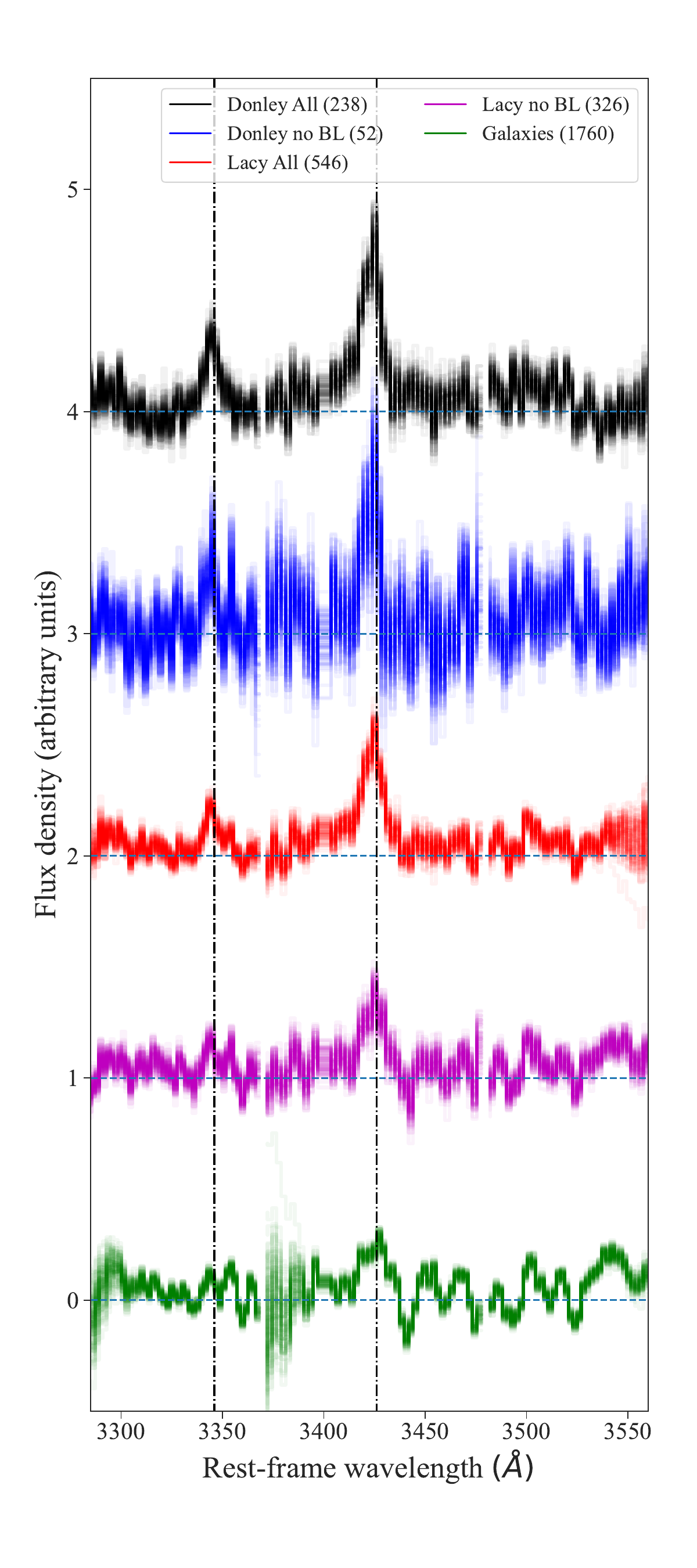}
\caption{Mean spectral stacks near the [\ion{Ne}{5}]$\lambda$3426 line region. The 88 AGNs that have individual [\ion{Ne}{5}]$\lambda$3426 detections are not included. The red and black lines are stackings of AGN candidates selected by the Lacy and Donley wedges, respectively. The magenta and blue lines are the corresponding stacks after removal of objects with broad line features. A stack of 24$\,\mu$m selected sources outside the Lacy wedge is shown in green. The number of spectra entering each stack are labeled in the legend. \label{stacknev}}
\end{figure}

In this section we stack spectra to search for [\ion{Ne}{5}]$\lambda$3426 signals in the redshift range 0.13$<$z$<$1.40 for all of our sources.
Firstly, all of the spectra are shifted to rest-frame and normalized to the same noise level, calculated according to Equation~\ref{eq2}, to ensure balanced weights. 
A linear continuum is then subtracted from each spectrum before the residual spectra are stacked to obtain the mean spectrum. 
Results from 100 bootstrap realizations of the stacking procedure are shown in Figure~\ref{stacknev}.
The black and red lines show the stackings of all [\ion{Ne}{5}]$\lambda$3426 undetected spectra in the Donley and Lacy wedges, respectively. 
The [\ion{Ne}{5}]$\lambda$3346,3426 doublet is clearly detected with mean SNR~$>10$.
After excluding the broad line AGNs, 
the stacking of Donley candidates shows that the mean SNR is 1.7 in the [\ion{Ne}{5}]$\lambda$3426 line region (blue line in Figure~\ref{stacknev}). 
The reliability of the latter stacking is limited by the small sample size (only 54 objects are included).
The magenta line shows that the stacking of Lacy candidates without broad lines does feature [\ion{Ne}{5}]$\lambda$3426 emission (SNR$\sim$2.6). 
Finally, the spectral stack of objects outside Lacy's wedge (shown as the green line) does not show a clear detection (the SNR is only 1.7).
This indicates a lack of AGN activity. 
In summary, our stacking results confirm that AGN activity exists in the MIR color-selected AGN candidates, although their low [\ion{Ne}{5}]$\lambda$3426 SNR implies that the emission line cannot be detected in every of the individual optical spectra.

\section{AGN decomposition in the MIR}\label{dec}

Although the stacked spectra of wedge-selected AGN candidates do reveal [\ion{Ne}{5}]$\lambda$3426 signals, confirming their AGN activity, 
it is still unclear what the quantitative level of AGN activity in individual galaxies is. 
In this section,  we develop a novel method to decompose the AGN and star formation components in the MIR. 
Separating the AGN emission from the host galaxy is a crucial issue in studies of the AGN-galaxy connection, particularly for Type I AGNs \citep{2015A&A...576A..10C}. 
Several attempts have been made to construct AGN models by solving the radiative transfer equation \citep{2006MNRAS.366..767F,2012MNRAS.420.2756S,2016MNRAS.458.2288S}. 
Multiwavelength SED fitting codes that utilize these models have been developed \citep[e.g.,][]{2019A&A...622A.103B,2020MNRAS.491..740Y}. 
However, these methods often require high SNR photometry across all wavelengths, which can be challenging to obtain, particularly in the far-infrared (FIR) regime. 
Given our lack of high quality data in the FIR, MIR colors are adopted.

\begin{figure}[ht!]
\epsscale{1.2}
\plotone{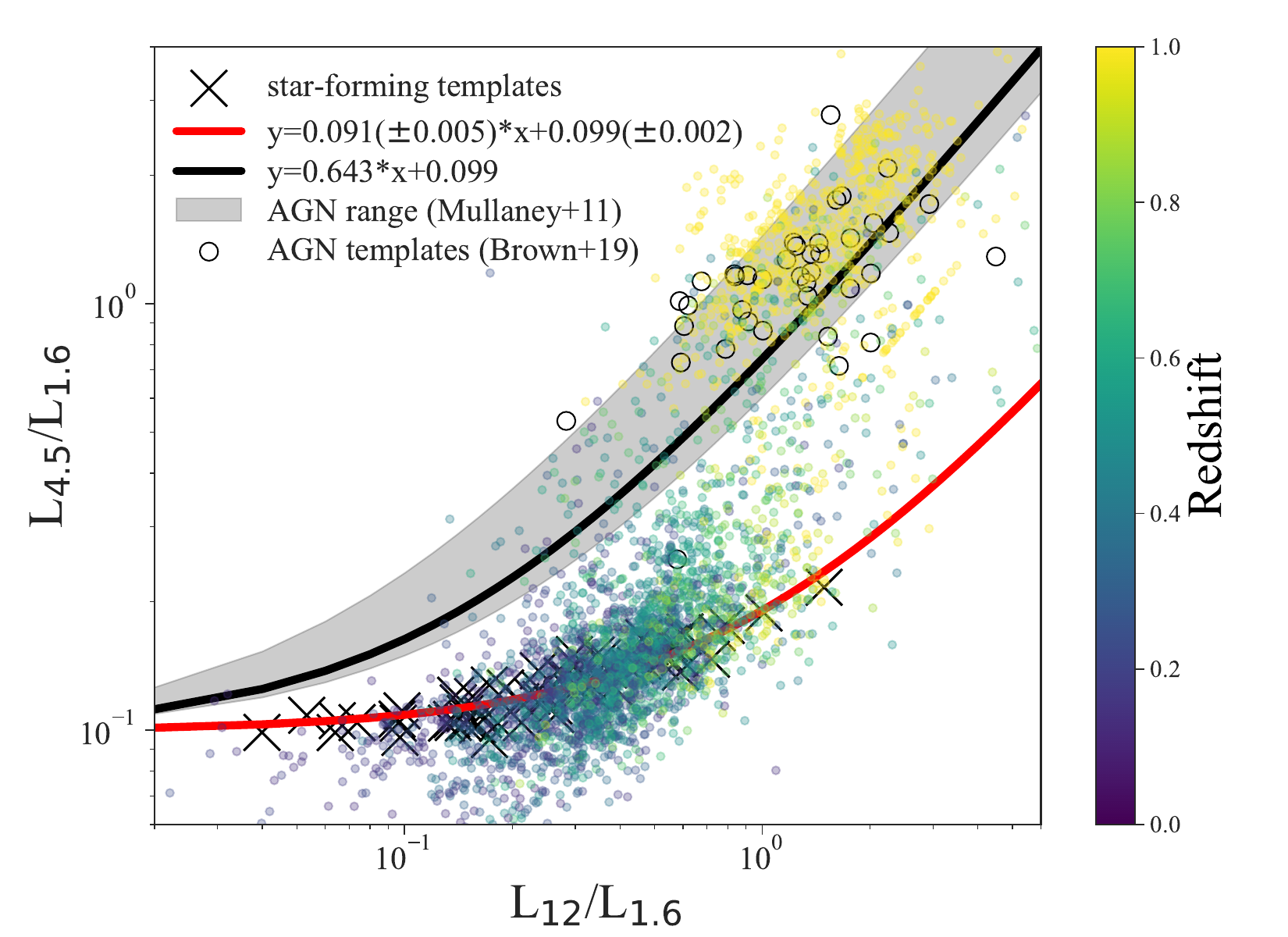}
\caption{L$_{12}$ vs L$_{4.5}$. Parameters are normalized to L$_{1.6}$ to avoid correlation caused by redshift. Black crosses are the star forming templates from \citep{2021ApJ...912..161H}, with the red line showing a linear fit to them. The color bar indicates redshift. The black open circles are AGN templates from \citet{2019MNRAS.489.3351B}. The grey-shaded region shows the range of $\alpha$ values from \citet{2011MNRAS.414.1082M}. In this figure, we only include 3679 objects ($\sim$91\% of the whole spectroscopic sample) that have photometry in at least three bands from 3.6 to 24$\mu$m. \label{l12_l45_fig}}
\end{figure}

We firstly apply a K-correction to calculate, for each source, three monochromatic luminosities at rest-frame wavelengths of 1.6$\mu$m, 4.5$\mu$m and 12$\mu$m, denoted as $L_{1.6}$, $L_{4.5}$ and $L_{12}$, respectively.  We base our K-corrections on the best-fitting MIR SED template, where this best-fitting template was identified by contrasting the observed 3.6 - 24$\mu$m photometry to each of the (redshifted and normalized) templates.
To this end, we employ as template library an atlas of UV-MIR SEDs for nearby galaxies by \citet{2014ApJS..212...18B}, which comprises 129 SEDs and covers diverse galaxy types.
We further complement the SED library with 113 AGN templates from \citet{2019MNRAS.489.3351B}.
We take the nearest observed-frame fluxes as input to the K correction procedure. For example, if an object has a redshift of z = 0.5, the rest 4.5 $\mu$m emission is redshifted to 6.75 $\mu$m, in which case we use observed 5.8 $\mu$m flux (nearest to 6.75 $\mu$m) to K-correct to get rest $L_{4.5}$.
In practice, across the variety of sources represented in our sample, the contributions from AGN and non-AGN components to the MIR luminosities will vary.  
The latter in principle can comprise both direct emission from stars (taken to dominate the 1.6 $\mu$m emission) and dust-reprocessed emission from star formation (more important at longer wavelengths).  
In an effort to disentangle these different components, we adopt the following decomposition method. 
For rest-frame $L_{4.5}$, the emission could originate from AGN-heated dust, stellar emission, and star-forming emission. 
For $L_{12}$, the stellar emission is expected to be negligible. 
In our analysis, we treat the derived $L_{4.5}$ and $L_{12}$ as the sum of AGN and non-AGN components,

\begin{equation}
L_{4.5}  = L_{4.5}^{*} + L_{4.5}^{AGN}, \label{eq5}
\end{equation}

and 

\begin{equation}
L_{12} = L_{12}^{*} + L_{12}^{AGN}, \label{eq6}
\end{equation}

We assume that the AGN emission in the MIR follows a power law distribution, which is an approach commonly used in the literature to describe the continuum emission of AGNs in this wavelength range \citep[e.g.,][]{2011MNRAS.414.1082M,2012MNRAS.425.3094C,2021MNRAS.503.2598B},

\begin{equation}
L_{12}^{AGN} = (12/4.5)^{\alpha} L_{4.5}^{AGN},  \label{eq7}
\end{equation}
 
\citet{2021ApJ...912..161H}  performed further classification of the SED atlas by \citet{2014ApJS..212...18B} into 5 populations based on MIR color and the EW of a polycyclic aromatic hydrocarbon (PAH) feature. The resulting 5 populations were labelled AGN, star forming, composite, quiescent, and blue compact.
We derive the correlation between $L_{4.5}^{*}$ and $L_{12}^{*}$ using their 65 star forming templates (shown as the red line in Figure\ref{l12_l45_fig}):

\begin{equation}
L_{4.5}^{*} =0.091(\pm0.005)L_{12}^{*}+0.099(\pm0.002)L_{1.6}. \label{eq8}
\end{equation}

After combining Equations~\ref{eq5} to~\ref{eq8}, we calculate L$_{4.5}^{AGN}$  as:

\begin{equation}
L_{4.5}^{AGN}= k(L_{4.5}-0.091L_{12}-0.099L_{1.6}),  \label{eq9}
\end{equation}

where $k$ equals $(1-0.091(12/4.5)^{\alpha})^{-1}$. 
Similarly, the star formation contribution at 12$\,\mu$m (L$_{12}^{*}$) is given by

\begin{equation}
L_{12}^{*}=k(L_{12}-(12/4.5)^{\alpha}L_{4.5}+0.099(12/4.5)^{\alpha}L_{1.6}). \label{eq12}
\end{equation}

All of the luminosities mentioned above can be determined once an appropriate value of $\alpha$ is adopted, parametrizing the slope of the AGN MIR SED.
\citet{2012ApJ...753...33D} constructed the median quasar SED based on the same 24 $\mu$m bright sample, and they found that $\alpha$ equals 0.45 and $k$ equals 1.16. 
Other studies, such as those by \citet{2011MNRAS.414.1082M}, show that the AGN SEDs have $\alpha$ values between -0.3 and 0.7, which translates to $k$ values between 1.07 and 1.22. 
Here we use $\alpha$ equals 0.45 to calculate the luminosities mentioned above, for consistency with \citet{2012ApJ...753...33D} whose sample is comprised within ours.  
We also consider $\alpha$ values between -0.3 and 0.7 to calculate the lower and upper limits to AGN luminosity for each object, respectively. 
As shown in the Figure~\ref{l12_l45_fig}, the AGN templates from \citet{2019MNRAS.489.3351B} are consistent with the $\alpha$ range between -0.3 and 0.7. This result reinforces our determination of $\alpha$ values.

Equation~\ref{eq12} would result in a negative value when L$_{12}$/L$_{1.6} < $(12/4.5)$^{\alpha}$(L$_{4.5}$/L$_{1.6}$-0.099). 
That would convey an AGN luminosity at 12 $\mu$m exceeding the total 12 $\mu$m luminosity, which is unphysical.  
We therefore consider such extreme objects as having their 12 $\mu$m emission entirely dominated by the AGN.  
Along a similar vein, the objects located below the red line are treated as pure star-forming galaxies.
This approach enables us to calculate the AGN fraction in a single band for every object. 

\begin{figure}[ht!]
\epsscale{1.15}
\plotone{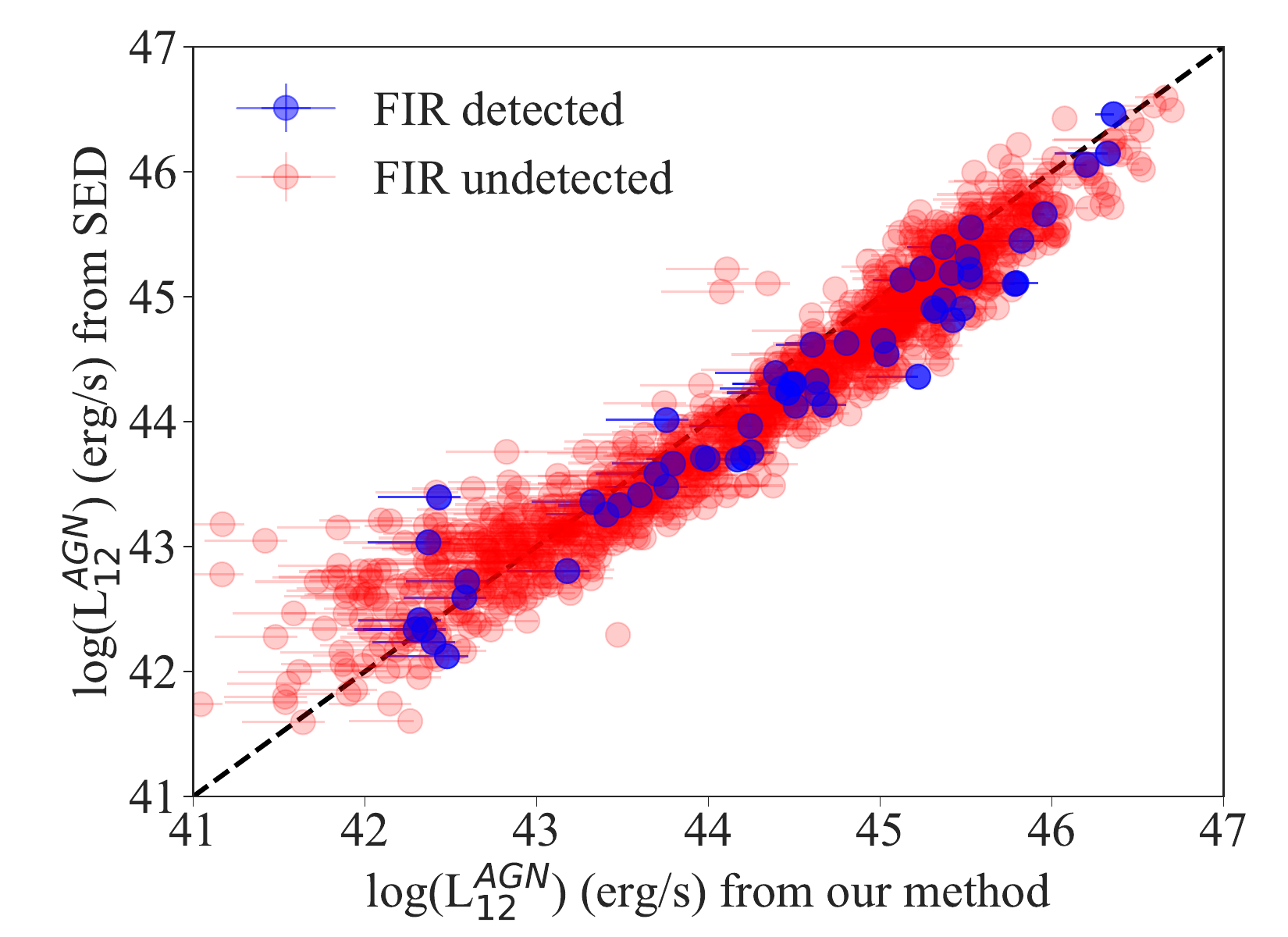}
\caption{Comparison of 12$\mu$m AGN luminosities derived from our method and those from SED fitting. The red dots are FIR undetected objects and the blue dots are FIR detected objects. The black dashed line shows 1:1 correlation. A good consistency is found between these two methods for AGN decomposition.} \label{agn2}
\end{figure}

To verify our method, we also do multiwavelength SED fitting for our objects from the optical to the FIR wavebands.
The photometries we used are listed in the Table~\ref{tab:data}.
The fitting is done by CIGALE code \citep{2019A&A...622A.103B,2020MNRAS.491..740Y}, which includes a AGN component in the fitting.
We compare the L$_{12}^{AGN}$ from our method and that from multiwavelength SED fitting, see Figure~\ref{agn2}. 
We label the obejcts that have SNR $>$ 3 in at least 3 FIR bands from 100 to 500 $\mu$m as ``FIR detected". 
The label indicates the objects with reliable FIR fitting.
As shown in the Figure~\ref{agn2}, we find the consistency between our method and SED fitting for both ``FIR detected" and ``FIR undetected" subsamples.
We further investigate the reliability of SF luminosity that we derived, see Section~\ref{sfrc}.

\begin{figure*}[ht!]
\epsscale{1.15}
\plottwo{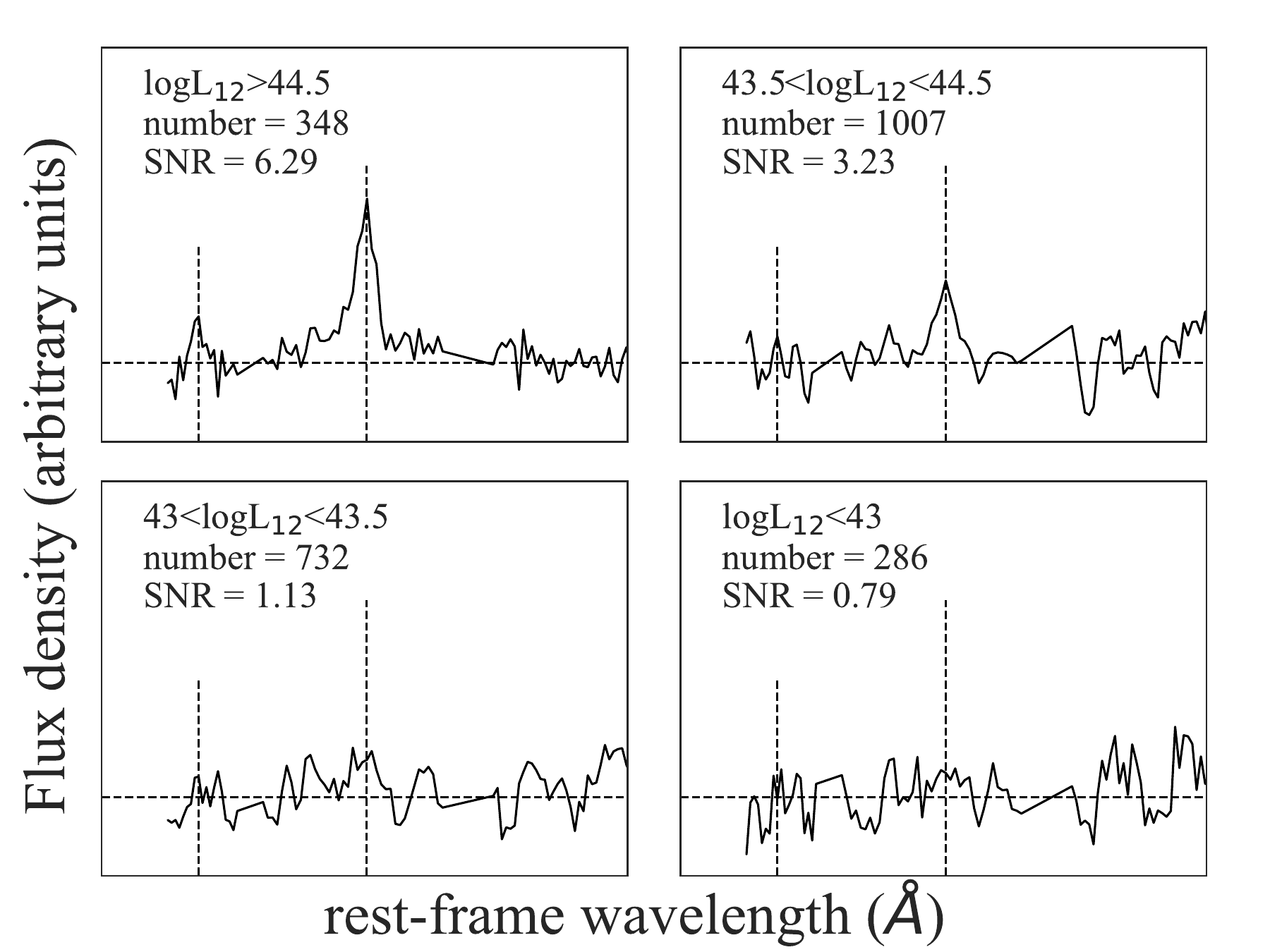}
{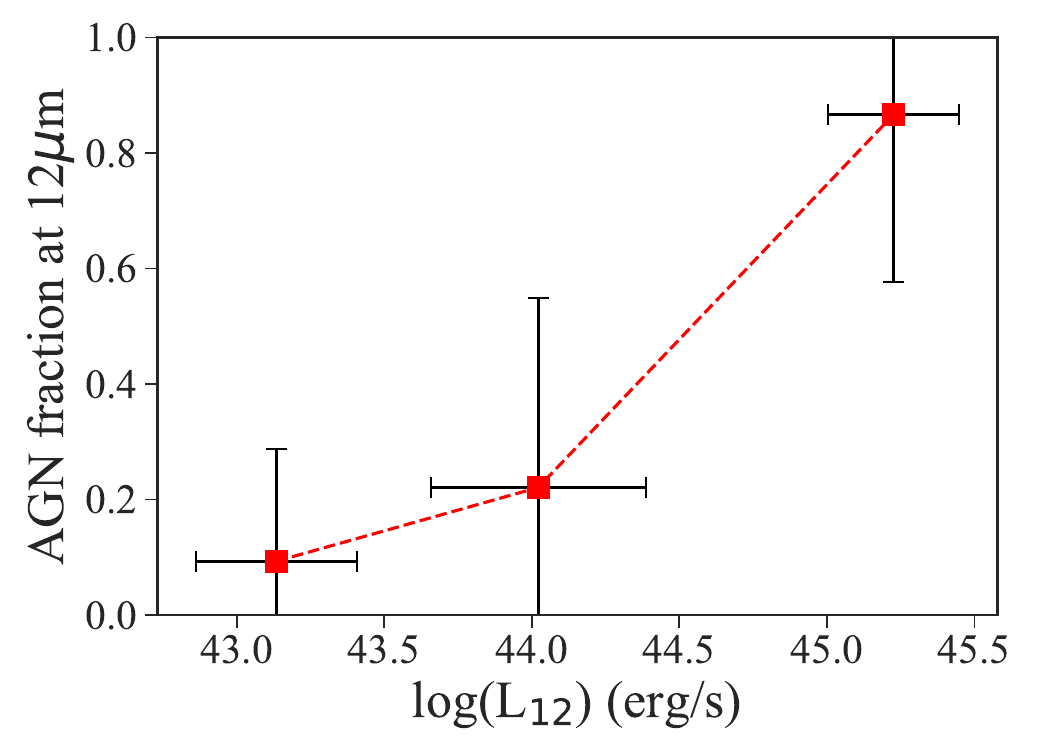}
\caption{\textit{Left Panel}: Mean spectral stacks of the [\ion{Ne}{5}]$\lambda$3426 line region for different 12$\mu$m total luminosities. Luminosity ranges, number of spectra and stacked SNR in each bin are marked in each panel. More luminous objects feature the stronger [\ion{Ne}{5}]$\lambda$3426 emission. \textit{Right Panel}: AGN fraction at 12 $\mu$m versus 12$\mu$m total luminosity. More luminous objects tend to have higher AGN fractions.} \label{nev_l12}
\end{figure*}

With the monochromatic rest-frame luminosities in hand, we first construct and investigate [\ion{Ne}{5}]$\lambda$3426 stacks for subsets of MIR-selected sources binned by their L$_{12}$ luminosity (see left panel of Figure~\ref{nev_l12}).  
We find that with greater L$_{12}$ luminosity comes stronger [\ion{Ne}{5}]$\lambda$3426 emission. 
The spectra of AGNs with L$_{12}> 10^{43.5}$ erg/s (two upper panels) display distinct [\ion{Ne}{5}]$\lambda$3426 emission. 
However, for mid-to-low luminosity objects (L$_{12} < 10^{43.5}$ erg/s), the SNRs of [\ion{Ne}{5}]$\lambda$3426 are lower than 2. 
This finding is consistent with previous studies on [\ion{Ne}{5}]$\lambda$14.3$\mu$m based on \textit{Spitzer}/IRS spectra. 
Specifically, \citet{2008ApJ...674L...9D} show that [\ion{Ne}{5}]$\lambda$14.3$\mu$m emission is stronger in Type I AGNs with higher luminosities. 
In our work, we did not differentiate between the Type I/II nature of AGNs and find the same result.  
Besides, in the right panel of Figure~\ref{nev_l12}, we plot AGN fraction at 12 $\mu$m against 12$\mu$m total luminosity. 
The AGN fraction increases along with the increasing 12$\mu$m luminosity.
These result from Figure~\ref{nev_l12} demonstrate that more luminous MIR sources tend to host more prominent AGN activity.

\begin{figure*}[ht!]
\epsscale{1.2}
\plotone{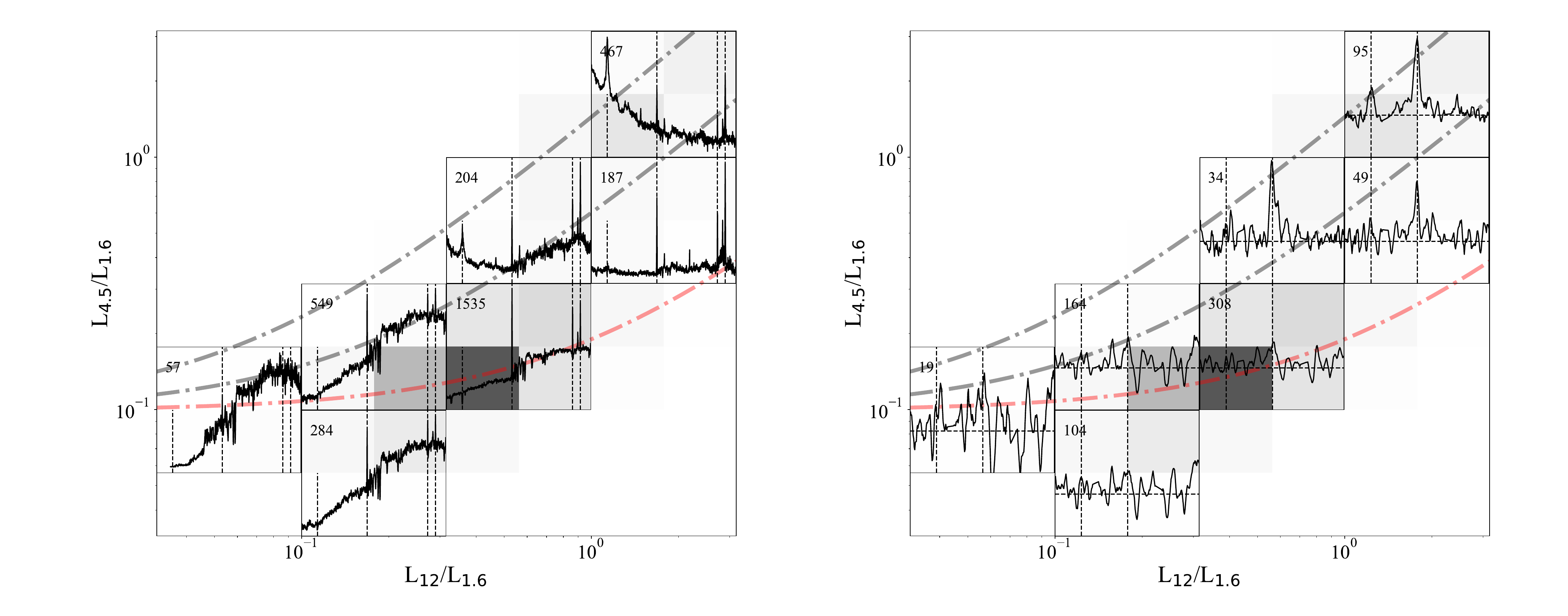}
\caption{\textit{Left panel}: Composite spectra across the L$_{12}$/L$_{1.6}$ versus L$_{4.5}$/L$_{1.6}$ diagram. No redshift limit was applied in this panel.  The rest-wavelength range of every inset panel is 2500-5200$\AA$. The number of combined spectra is marked. Besides, we mark KEx classification of each spectra in the upper left corner of each inset panel. The dashed lines mark \ion{Mg}{2}$\lambda$2800, [\ion{O}{2}]$\lambda$3727, H$\beta$ and [\ion{O}{3}$\lambda$5007 spectral lines from left to right, respectively. \textit{Right panel}: Mean stacking of the [\ion{Ne}{5}]$\lambda$3426 line region. In this panel, we only included spectra with $0.13<z<1.40$. The wavelength range of every inset panel is 3300-3550$\AA$. Two vertical dashed lines show the location of the [\ion{Ne}{5}]$\lambda$3426,3426 doublet. The number of spectra used in the stacking and the SNR of stacked [\ion{Ne}{5}]$\lambda$3426 is marked in the upper left corner of each inset panel.
The red dash-dotted lines in both panels show pure star formation relations while the The black dash-dotted show the range of $\alpha$ values we adopt. }\label{1245stacking}
\end{figure*}

We also  generate composite spectra based on the positions of the sources in Figure~\ref{l12_l45_fig}, following the same procedures described in Section~\ref{composite}.
The results are presented in Figure~\ref{1245stacking}.
Every inset panel in this figure corresponds to a bin of 0.5 dex in L$_{12}$/L$_{1.6}$ and 0.5 dex in L$_{4.5}$/L$_{1.6}$.
Type I quasar features (e.g., broad \ion{Mg}{2} lines) are clearly visible in the ($10^{0},10^{0}$) and ($10^{-0.5},10^{-0.5}$) bins.
However, spectra in the ($10^{-0.5},10^{-0.5}$) bin have a redder continuum, indicating a transitional state between Type I and Type II.
Spectra in the ($10^{0},10^{-0.5}$) bin exhibit Type II quasar features (narrow permitted lines and flat continua).
Consistently, [\ion{Ne}{5}]$\lambda$3426 stacking reveals detections in all these three bins.
Comparing these three composite spectra, we find that Type II AGNs usually have lower L$_{4.5}$  and higher L$_{12}$ compared with Type I AGNs, normalized by L$_{1.6}$.
This conclusion suggests that Type I AGNs are usually bluer and Type II are redder ("rising") even in the MIR regime.
Previous studies arrived at consistent conclusions \citep[e.g.,][]{2021ApJ...906...84O}.
By definition, Type II AGNs are obscured objects, by dust and gas from either the circumnuclear region or the interstellar medium (ISM) on the galactic scale. 
If the obscuration is from the ISM, the extinction ( typically A$_{V} \sim$0-2 mag, see e.g., \citealt{2005MNRAS.358..363C}) would result in a redder optical-NIR SED. 
However, it should not affect the MIR SED since the obscuring material is optically thin in the MIR \citep{2012ApJ...748..142D}. 

On the other hand, central obscuration near the nucleus could also
result in the observed difference in the SEDs of Type I and II AGNs. 
Nevertheless, the circumnuclear extinction may not explain the difference of MIR SED shapes in the majority of AGNs  (log$N_{H}\sim$20-24 cm$^{-2}$),
as it was found that the SEDs do not change significantly across a wide range of circumnuclear extinction, with log$N_{H}$ values from 20 to 24 cm$^{-2}$ \citep[see Figure 15 in][]{2021MNRAS.503.2598B}, even after correcting for the silicate absorption at 9.7$\mu$m that dominates the ISM \citep[e.g., ][]{2012ApJ...755....5G}. 
As an alternative explanation,  \citet{2021MNRAS.503.2598B} argued that the  ``intrinsic" rising SED may arise from complex nuclear dusty structures such as extended polar dust \citep[see][]{2021ApJ...906...84O}.  
In these heavily obscured systems (e.g., Compton-thick systems with log$N_{H}>24$ cm$^{-2}$) , the hot dust emission from the nuclear region is partly absorbed by the extended dust components and re-emitted at longer wavelengths, leading to  a cooler (and thus redder in the MIR) SED.

\section{AGN and star formation}\label{pop}

\subsection{AGN bolometric luminosity }\label{bol}
In this study, we use the monochromatic calibration described in \citet{2012MNRAS.426.2677R}:

\begin{equation}
\log(L_{bol}^{AGN})=0.822 (\pm0.096)\ \log(L_{12}^{AGN})+8.915 (\pm4.303),  \label{eq13}
\end{equation}

to convert L$_{12}^{AGN}$ into AGN bolometric luminosity.

\begin{figure}[ht!]
\epsscale{1.15}
\plotone{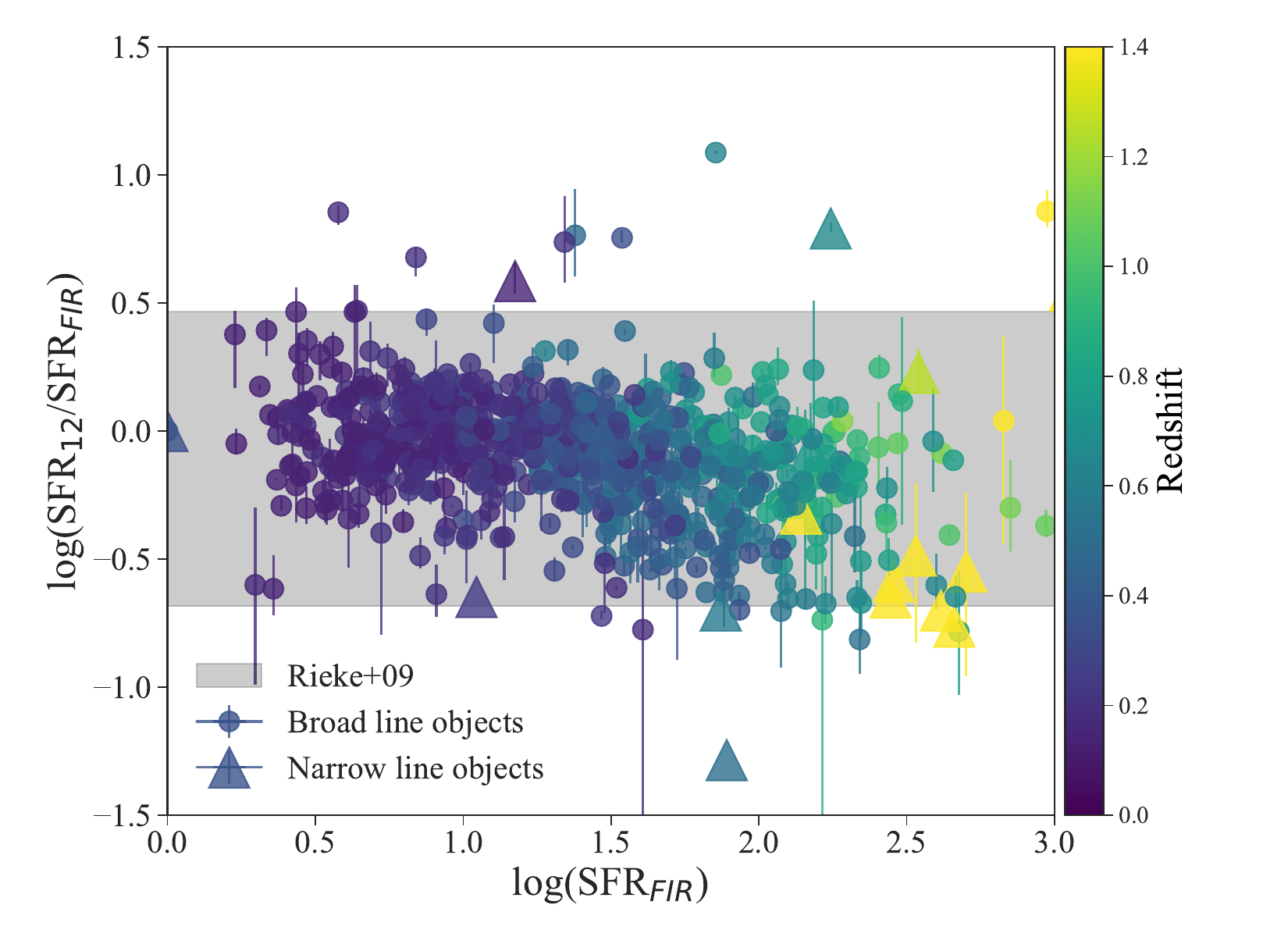}
\caption{Comparison of SFRs derived from 12$\mu$m and those from FIR emission. Triangles mark broad line objects and dots are narrow line objects. The grey-shaded shows the template predication of SEDs from \citet{2009ApJ...692..556R}.} \label{sfr2}
\end{figure}

\subsection{Star formation rate}\label{sfrc}

The conversion of MIR monochromatic luminosity to star formation rate (SFR) has been explored in several studies \citep[e.g.,][]{2007ApJ...666..870C,2009ApJ...692..556R}.
In this study, we use the calibration presented in \citet{2013ApJ...774...62L} to estimate SFR from 12$\mu$m luminosity, in units of M$_{\odot}$/yr:

\begin{equation}
\log(SFR_{12}) = 1.03\ \log(L_{12}^{*})-43.77. \label{eq10} 
\end{equation}

We also obtain FIR luminosities ( the wavelength range in this work is defined as 40-500 $\mu$m) by fitting the templates from \citet{2001ApJ...556..562C} to approximately 20\% of our objects 
(those that have SNR $>$ 3 in at least 3 FIR bands from 100 to 500 $\mu$m, see Table~\ref{tab:data}).
We then convert the FIR luminosities to SFRs using:

\begin{equation}
\log(SFR_{FIR}) = 1.02\ \log(L_{FIR})-44.42. \label{eq11} 
\end{equation}

This equation is originally from \citet{1998ARA&A..36..189K}.
We find a good agreement between the SFRs derived from the 12$\mu$m luminosity and the FIR luminosity, see Figure~\ref{sfr2}. 
This result is also consistent with the expectations from star-forming templates \citep{2009ApJ...692..556R}, which further confirms our MIR-based SFR estimate.

\begin{figure}[ht!]
\epsscale{1.15}
\plotone{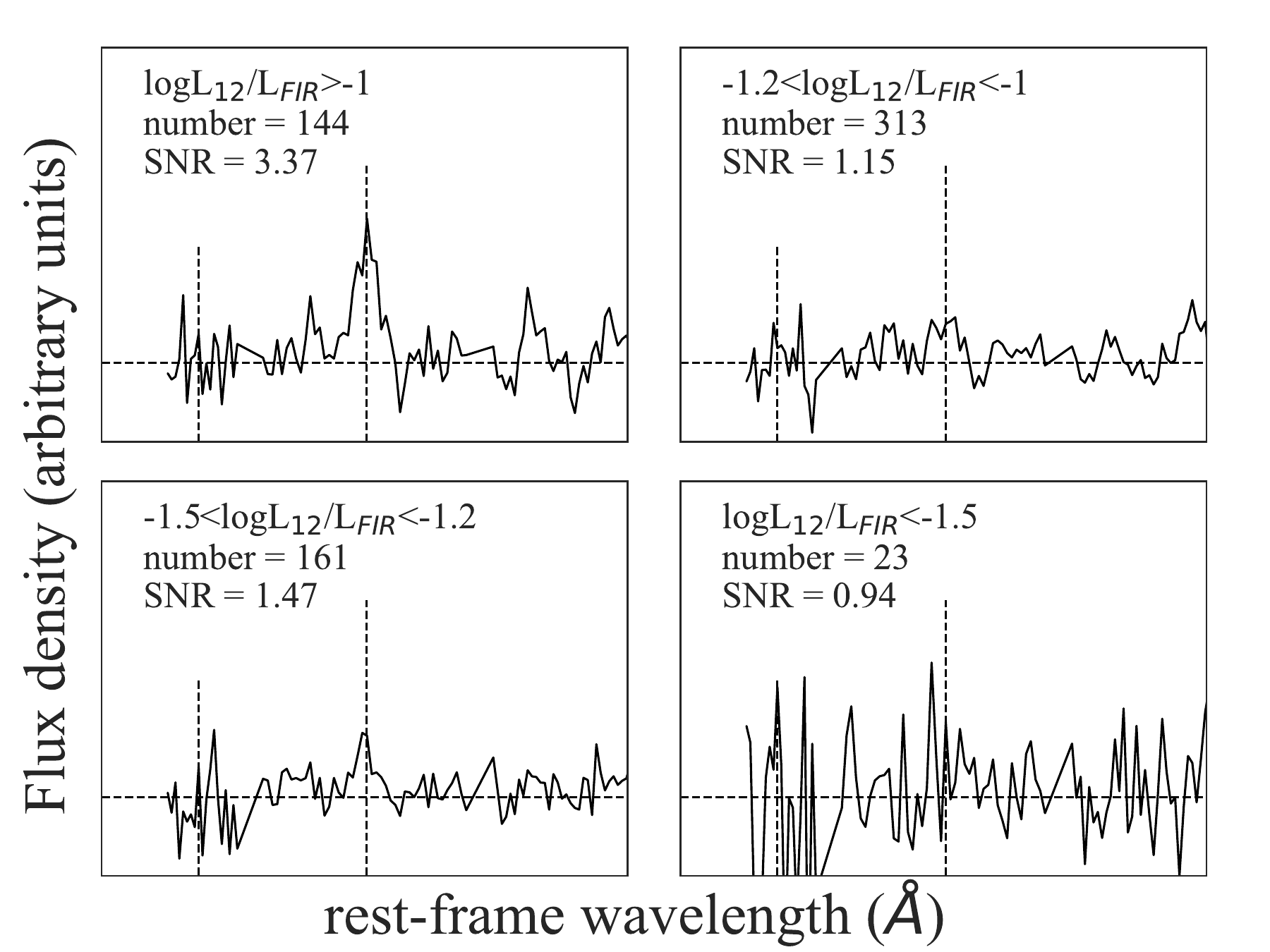}
\caption{Mean spectral stacks of the [\ion{Ne}{5}]$\lambda$3426 line region for different L$_{12}$/L$_{FIR}$ ratios. The range of the considered luminosity ratio, number of spectra and stacked SNR in each bin are marked in each panel.  Objects which are more luminous at 12$\mu$m compared with the FIR have the stronger [\ion{Ne}{5}]$\lambda$3426 emission.}\label{12f_lbol}
\end{figure}

As a sanity check, we also stacked the [\ion{Ne}{5}]$\lambda$3426 spectra based on the ratio between the total 12$\mu$m luminosity and their FIR luminosity.  Figure~\ref{12f_lbol}, the objects featuring the strongest MIR emission relative to their FIR output tend to be associated with a stronger [Ne V] signal, indicative of their MIR excess being due to stronger AGN activity.

\subsection{Correlation between AGN and star formation}\label{A41}

The relation between AGN luminosity and star formation rate has been extensively studied \citep{2012A&A...545A..45R,2015ApJ...806..187A,2017MNRAS.466.3161S,2018MNRAS.478.4238D,2020MNRAS.498.2323J}. 
Theoretical models suggest that the star formation activities and AGN activities in galaxies may be related \citep{2005Natur.433..604D,2006ApJS..163....1H}.
Gas infall can trigger both AGN and star formation \citep{2008MNRAS.391..481S}, while AGN feedback can suppress star formation in the host galaxy \citep{2005MNRAS.361..776S}.
Observational studies have shown a positive relationship between AGN and star formation in some cases \citep{2008ApJ...684..853L,2013ApJ...773....3C,2018MNRAS.478.4238D}, while in other studies, the relationship becomes weak or absent \citep{2010A&A...518L..26S,2012ApJ...760L..15H,2015ApJ...801...87B,2016MNRAS.460..902B,2017MNRAS.466.3161S}.
\citet{2010A&A...518L..26S} proposed a two-phase model where, for low luminosity AGNs, the AGN-SF relation is weak, reflecting a secular evolutionary path, whereas for high luminosity AGNs, the AGN and star formation in the host galaxy are linked due to the merging process, boosting both.
The relationship between AGN and star formation is further complicated by the different timescales of AGN and star formation activities.
\citet{2014ApJ...782....9H} considered the variability of AGNs and found a consistent AGN-SF relation with the two-phase model proposed by \citet{2010A&A...518L..26S}.

\begin{figure*}[ht!]
\epsscale{1.15}
\plottwo{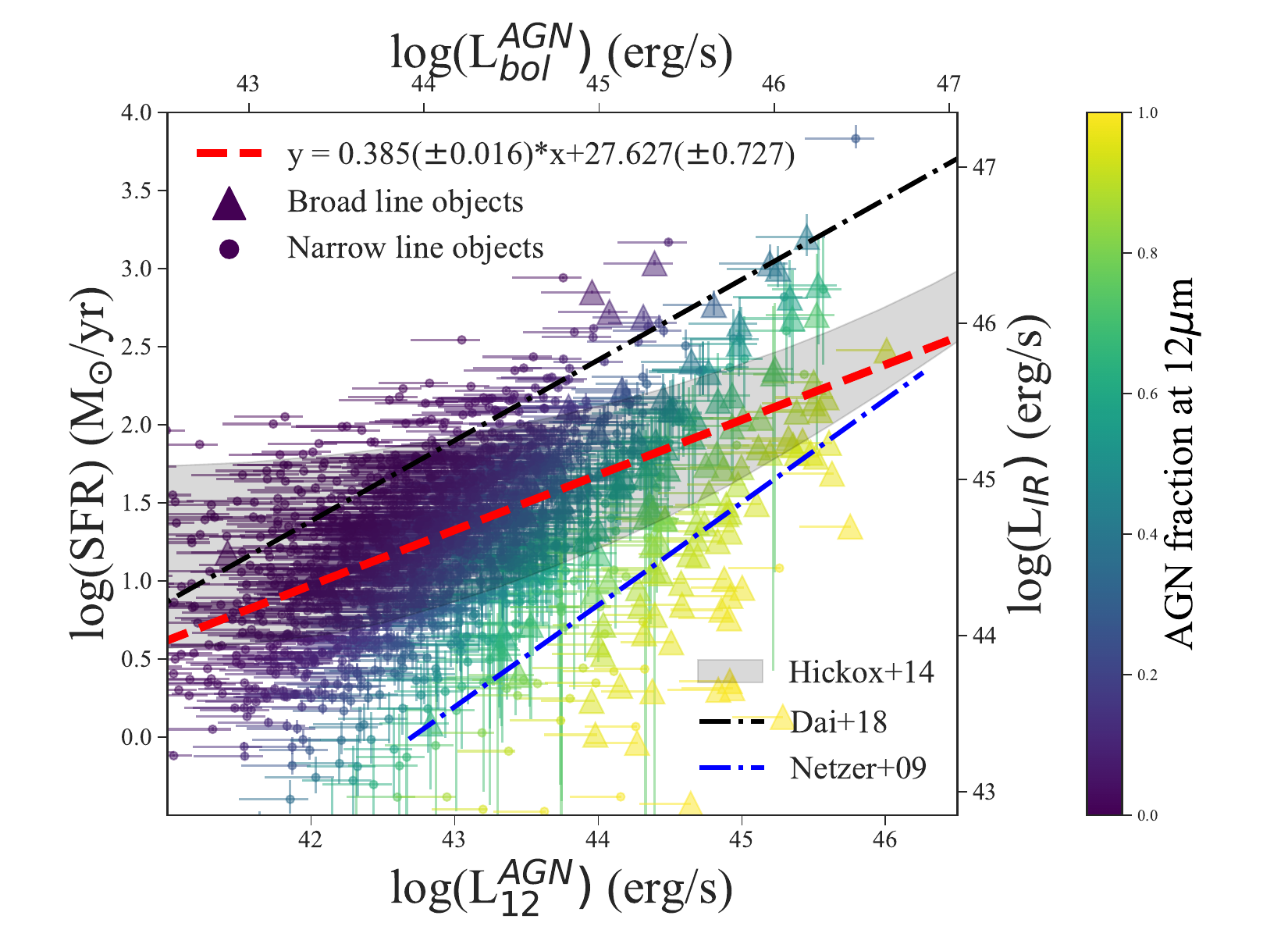}{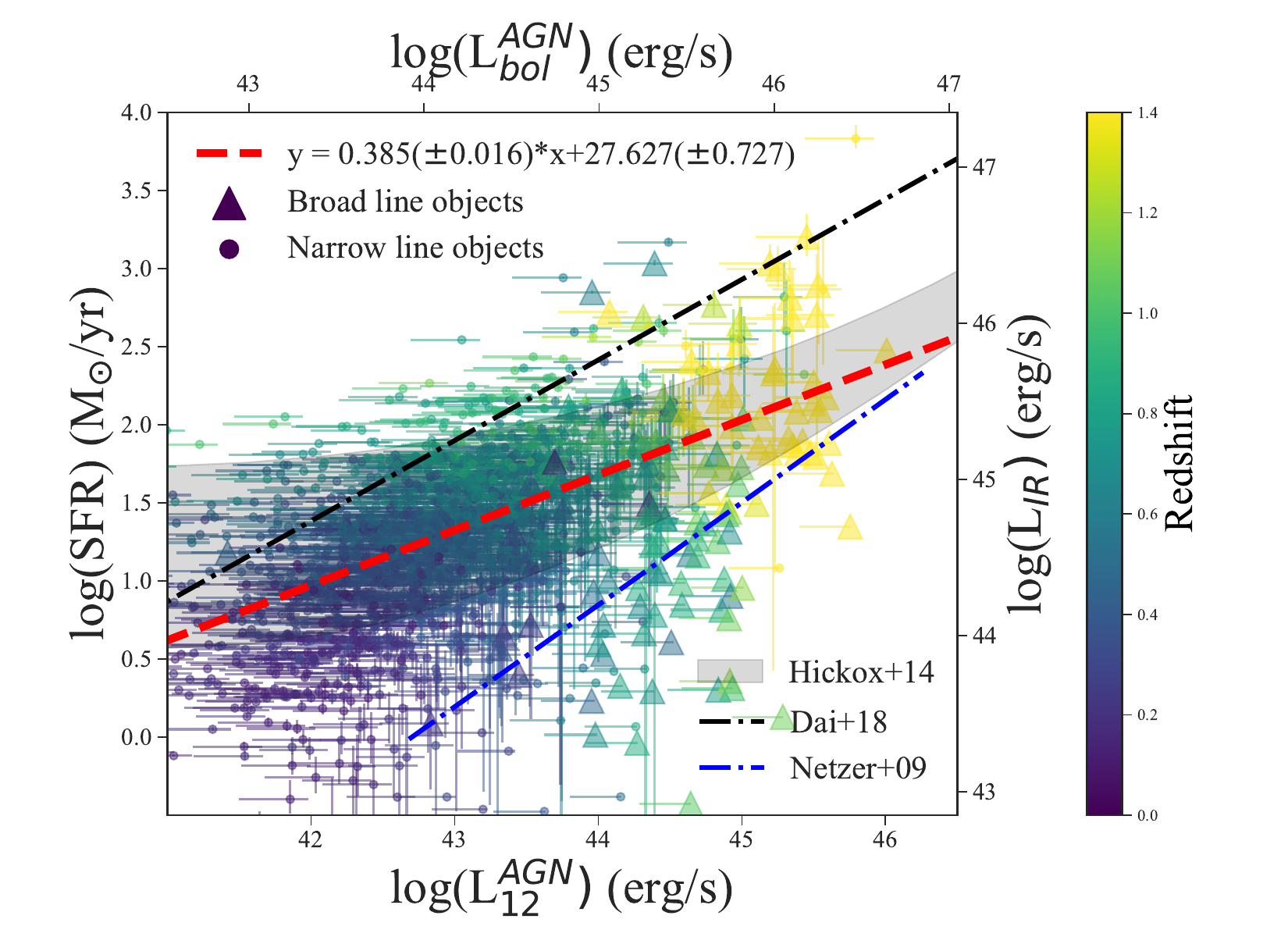}
\caption{SFR versus AGN 12\,$\mu$m luminosity. Colors are AGN fraction at 12$\mu$m and redshift in the left and right panels, respectively. Triangles mark broad line objects and dots are narrow line objects. The red dashed line is a linear fit to all objects with SFR and AGN luminosity measurements (1811 objects, see Section~\ref{dec}). Fitting parameters are listed in the upper left legend. The blue dash-dotted line is the relation found by \citet{2009MNRAS.399.1907N}. The black dash-dotted line is the relation derived by \citet{2018MNRAS.478.4238D}. The grey polygon shows the model prediction from \citet{2014ApJ...782....9H}. The bolometric luminosities are converted from L$_{12}^{AGN}$ using the correction described in \citet{2012MNRAS.426.2677R}.}\label{sfr_lbol}
\end{figure*}

Figure~\ref{sfr_lbol} shows the SFR as a function of AGN luminosity, 
along with the relations from  \citet{2009MNRAS.399.1907N}, \citet{2014ApJ...782....9H}, and \citet{2018MNRAS.478.4238D}.
The red dashed line represents the linear fit to our parent spectroscopic sample  
showing a positive relation between AGN and star formation activities.
We note that only 1811 objects with measurements of both L$_{bol}^{AGN}$ and SFR are included, and objects below the red line (i.e. pure star-forming galaxies) 
and above the black line (i.e. pure AGNs) in Figure~\ref{l12_l45_fig} are excluded.
Nevertheless, a larger sample of less luminous AGNs with logL$_{bol}^{AGN}<$~43 erg/s is required to confirm the possible flattening of the AGN-SF relation at the low luminosity end.
It is worth noting that the sample selection could affect the linear relation shown in Figure~\ref{sfr_lbol}. 
For example, \citet{2018MNRAS.478.4238D} studied an FIR-bright quasar sample at similar redshifts, 
while \citet{2009MNRAS.399.1907N} focused on a local SDSS AGN sample.
These differences could explain that our sample lies in between the two referenced relations. 
The lower fraction of FIR-bright objects could explain our lower SFR than \citet{2018MNRAS.478.4238D},
while the higher redshift in our sample, as well as the 24$\mu$m selection result in the higher SFR compared to \citet{2009MNRAS.399.1907N}. 
As a cautionary note, we emphasize that our sample comprises AGNs spanning a wide range in lookback times, and higher-redshift galaxies are well known to host larger gas reservoirs and feature higher SFRs.  
Therefore, population effects, stemming from the broad range in galaxy properties and redshifts included within the sample, may contribute to shaping the relation seen in Figure~\ref{sfr_lbol}. 
The presence of an L$_{bol}^{AGN}$ -- SFR correlation may thus not necessarily be taken as a direct imprint of individual galaxies experiencing temporal co-evolution. 
We illustrate the redshift distribution of sources across the L$_{bol}^{AGN}$ -- SFR diagram in the right panel of Figure~\ref{sfr_lbol}. 
At $L_{bol}^{AGN} < 10^{45}$ erg/s, higher redshift objects are seen to feature higher SFR.  This reflects the redshift evolution of the SF main sequence \citep[e.g.,]{2015A&A...575A..74S}.
The SFR shows a weaker dependency on the AGN luminosity for objects at similar redshift.
The lack of a strong L$_{bol}^{AGN}$ -- SFR relation once controlling for redshift is also reported by \citet{2015MNRAS.453..591S}, who apply an X-ray AGN selection and luminosity measurement.
Due to the variability of AGNs, the measurement of SFR and L$_{bol}^{AGN}$ is sensitive to the time scale of the relevant physical processes.
Short time-scale variations in such systems could flatten the relationship between SFR and L$_{bol}^{AGN}$ \citep{2015MNRAS.453..591S}.
Other factors such as location of SF activity could also wash out the relationship \citep{2015MNRAS.449.1470V}.
Broad line objects dominate the population with L$_{bol}^{AGN} > 10^{45}$ erg/s and they occupy the high-SFR and high-L$_{bol}^{AGN}$ space of Figure~\ref{sfr_lbol}.
They show a strong L$_{bol}^{AGN}$ -- SFR relation.
Based on X-ray data, \citet{2012A&A...545A..45R} found the same conclusion that the black hole activity tracks global SF of the host galaxy in high luminosity AGNs.
They argued that this connection could result from rapid gas inflow associated with mergers. 
The gas inflow fuels both AGN accretion and nuclear star-formation.
Thus the L$_{bol}^{AGN}$ -- SFR relation in our sample could stem partly from such interconnection, with additional contributions from the varying population properties across the range of redshifts explored.


\section{Conclusion}\label{sss}

We present an optical spectroscopic sample of 24$\mu$m bright objects in the Lockman Hole field from the SWIRE survey, obtained with MMT/Hectospec. Out of all targets with F$_{24}>$400$\mu$Jy and $17.7 < r < 22.5$, 
$\sim$54\% is covered in our sample. 
The sample is further supplemented with optically bright ($r < 17.7$) sources above the same 24$\mu$m threshold with spectra from SDSS.
Given the redshift distribution of our sample, we used [\ion{Ne}{5}]$\lambda$3426 as the preferred diagnostic to evaluate the presence of AGNs in the 4035 spectra with $z>0.13$, 
and identified 88 robust [\ion{Ne}{5}]$\lambda$3426-selected AGNs with SNR$>$3. 
We combined the sample with previously identified Type I broad-line AGNs, together yielding a total sample of 887 AGNs (844 of them are Type I AGNs).

Compared with various AGN MIR color selections, we confirm that 53\% of  objects selected by the Lacy color criteria are also spectrally confirmed AGNs. 
The percentage increases to 84\% for the Donley wedge. 

To determine the potential AGN activity within [\ion{Ne}{5}]$\lambda$3426 undetected, wedge-selected AGN candidates, we stack their spectra to look for [\ion{Ne}{5}]$\lambda$3426 emission.
We find that the stacked [\ion{Ne}{5}]$\lambda$3426 strength increases with increasing 12$\mu$m luminosity, indicating that the most MIR luminous objects tend to host more prominent AGN activity. 
This result is also consistent with the conclusion from our decomposition, which reveals that objects with greater 12$\mu$m luminosity have larger AGN fraction. 

The stacking of both Lacy and Donley candidates in the [\ion{Ne}{5}]$\lambda$3426 region
shows clear detections.
Our stacking analysis confirms that AGN activity is present among MIR color-selected AGN candidates, although their low SNR [\ion{Ne}{5}]$\lambda$3426 cannot be detected in all of the individual optical spectra.

To quantify the AGN and star forming contribution in each object, we develop a novel method to decompose the AGN and star formation components in the MIR.
The relation between SFR and AGN bolometric luminosity of our sample is consistent with the two-phase model prediction over a dynamic range in L$_{bol}^{AGN}$ of $10^{43}$ to $10^{47}$ erg/s. 
We discuss that part of this correlation is attributed to the higher-redshift objects in our sample populating the more luminous L$_{bol}^{AGN}$ and higher SFR regime. 

In summary, our study confirms spectroscopically the presence of AGN activity in MIR color-selected AGN candidates.  
The AGN strength is larger in the more luminous MIR sources.
We develop a novel method to decompose the AGN and star formation contribution using rest-frame MIR colors. 
Based on this method, we find a positive correlation between AGN luminosity and SFR.

~\\
The authors would like to thank Cheng Cheng, Hai Xu, Piaoran Liang, Yaru Shi, Shumei Wu,  Gabriel Oio and Xianzhong Zheng for helpful discussions. 
This work is sponsored by the National Key R\&D Program of China for grant No.\ 2022YFA1605300, 
the National Nature Science Foundation of China (NSFC) grants No. \  12273051 and 11933003.
The authors gratefully acknowledge support from the Royal Society International Exchanges Scheme (IES$\backslash$R1$\backslash$211140) and the Chinese Academy of Sciences President's International
Fellowship Initiative (grant no. 2022VMB0004). T. C. acknowledges the China Postdoctoral Science Foundation (Grant No.\ 2023M742929).
Additional support came from the Chinese
Academy of Sciences (CAS) through a grant to the South
America Center for Astronomy (CASSACA) in Santiago, Chile.


\bibliography{sample631}
\bibliographystyle{aasjournal}
\bibstyle{thesisstyle}


\end{CJK*}
\end{document}